\documentclass[11pt,english]{article}
\usepackage[T1]{fontenc}
\usepackage[latin9]{inputenc}
\usepackage{array}
\usepackage{amsmath}
\usepackage{amssymb}
\usepackage{graphicx}
\usepackage{esint}

\makeatletter

\providecommand{\tabularnewline}{\\}

\newcommand{\lyxaddress}[1]{
\par {\raggedright #1
\vspace{1.4em}
\noindent\par}
}

\usepackage{a4wide}
\usepackage{hyperref}
\numberwithin{equation}{section}

\makeatother

\usepackage{babel}
\begin{document}

\title{On form factors of boundary changing operators}

\author{Z. Bajnok$^{\alpha}$ and L. Hollo$^{\alpha,\beta}$}

\maketitle

\lyxaddress{\begin{center}
$\alpha$: \emph{MTA Lend\"ulet Holographic QFT Group, Wigner Research
Centre for Physics}\\
\emph{H-1525 Budapest 114, P.O.B. 49, Hungary}\\
$\beta$: \emph{Laboratoire de Physique Théorique, École Normale Supérieure,
}\\
\emph{24, rue Lhomond 75005 Paris, France}
\par\end{center}}
\begin{abstract}
We develop a form factor bootstrap program to determine the matrix
elements of local, boundary condition changing operators. We propose
axioms for these form factors and determine their solutions in the
free boson and Lee-Yang models. The sudden change in the boundary
condition, caused by an operator insertion, can be interpreted as
a local quench and the form factors provide the overlap of any state
before the quench with any outgoing state after the quench. 
\end{abstract}

\section{Introduction}

Integrable 1+1 dimensional systems are very special quantum field
theories as they can be solved exactly \cite{Mussardo:1281256,Bajnok:2013sa}.
The models and the obtained solutions are interesting in many respects.
First, they appear on various areas of theoretical physics ranging
from statistical physics to string theory. Second, the exact solutions
can be compared to and test alternative approximate solutions. 

The procedure of solving integrable theories consists of two steps.
In the first step the scattering ($S$) and reflection ($R$) matrices,
connecting asymptotic initial and final states, are determined. These
contain the on-shell information of a given bulk or boundary quantum
field theory. In the second step restrictive functional equations
are formulated for the form factors involving the already determined
$S$ and $R$ matrices. The solutions of these equations provide off-shell
information which then can be used to calculate the correlation functions
via the spectral representation. 

Recently there have been increasing interest in quench type problems.
They appear when, at a given time, a parameter of the physical system
is changed. They are relevant in statistical physics and solid state
problems. On the string theory side they appear when the strings split,
fuse or change their boundary conditions \cite{Lucietti:2003ki,Lucietti:2004wy}.
So far the integrable approaches assumed a squeezed coherent (boundary)
state form of the system after the quench, see \cite{Fioretto:2009yq,Sotiriadis:2013fca,Horvath:2015rya}
and references therein. Contrary, we would like to analyze a different
quench, which is related to form factors. As an example let us suppose
that we introduce a quench in a system at a moment by inserting a
local operator $\mathcal{O}$, which we can even integrate in space
$\int\mathcal{O}(x,0)dx$. In the quench framework we are interested
in how a given state (say the vacuum) will evolve after the quench.
This is probed by the matrix elements
\begin{equation}
\langle\theta_{1},\dots,\theta_{n}\vert\int\mathcal{O}(x,0)dx\vert0\rangle=F^{\mathcal{O}}(\bar{\theta}_{n},\dots,\bar{\theta}_{1})\delta_{P}\quad,\qquad\bar{\theta}=\theta+i\pi\,,
\end{equation}
which is basically the form factor of the operator $\mathcal{O}$,
and $\delta_{P}$ projects onto zero momentum states. Clearly, form
factors do not exponentiate, except from free theories. This quench
is, however, localized in time, and cannot be regarded as a change
of a parameter of the model. 

In the following we will be interested in another integrable quench,
which changes the parameters of the theory but still corresponds to
form factors. We analyze an integrable boundary system in which at
a moment we change the integrable boundary condition from $\alpha$
to $\beta$ by inserting a boundary condition changing operator. These
kinds of boundary quenches have been used to calculate the Loschmidt
echo in the Resonant Level Model \cite{Vasseur:2013rda}. As the vacuum
evolves to the form factors of the boundary condition changing operator
we formulate axioms to determine these quantities. 

In \cite{Lesage:1998hh,Lesage:1997tc} the authors proposed form factor
axioms both for boundary operators and for boundary changing operators.
First they adopted the boundary form factor axioms from lattice models
\cite{Jimbo:1994np} and adjusted them for the relativistic kinematics.
Then, on the example of the free massive fermion model they generalized
them for operators which change the boundary condition and they further
analyzed the solutions of these equations. Finally, they extended
the axioms for non-trivial bulk scatterings and investigated the sinh-Gordon
model, where they calculated the form factors of boundary changing
operators up to 4 particles. They also extended the analysis for massless
scatterings and applied the results for the double well problem of
dissipative quantum mechanics. 

In \cite{Bajnok:2006ze} the authors analyzed the form factors of
local boundary operators from a different perspective. They derived
a closed set of boundary form factor axioms from the boundary reduction
formula \cite{Bajnok:2002cm}. These axioms, besides of the previous
ones of \cite{Lesage:1998hh}, additionally contained the boundary
kinematical singularity axiom, making the whole system complete in
the sense, that the space of the solutions is in one to one correspondence
with the space of all local boundary operators of the UV boundary
conformal field theory \cite{Szots:2007jq}. This boundary form factor
program was carried out in many integrable models and was generalized
to nondiagonal scattering theories \cite{CastroAlvaredo:2006sh,CastroAlvaredo:2007pe,Takacs:2008je,Lencses:2011ab}. 

The aim of the present paper is to extend this form factor program
for boundary changing operators, i.e. our axioms, additionally to
the axioms of \cite{Lesage:1998hh}, contain the boundary changing
analogue of the boundary kinematical singularity axiom. We also show
that our axioms are complete in the above sense, as we find as many
solutions as many boundary changing local operators exists in the
UV limiting boundary conformal field theory.

The paper is organized as follows: In Section \ref{sec:Form_factor_axioms}
we introduce the theory of form factors in integrable field theories
and present our proposals for the boundary changing form factor axioms.
Various consistency checks are presented and we show the general method
to solve them. Their applicability to the calculation of two point
functions is also explained. In Section \ref{sec:Model_studies} we
solve the axioms in case of the free boson and Lee-Yang theories.
In the free boson theory direct field theoretical approach is also
presented. In case of the Lee-Yang model two-point functions of boundary
fields are calculated by summing up few particle form factor contributions
and compared, at short distance, to the conformal field theory prediction.
Their agreement is a solid confirmation of our form factor solutions.
We end the main part of the paper by the conclusion in Section \ref{sec:Conclusion}.
Some technical details are relegated to the two appendices. In Appendix
\ref{sec:ZF_algebra} a formal derivation of the axioms from the Zamolodchikov-Faddeev
algebra is shown. In Appendix \ref{sec:Neumann_to_Dirichlet} we study
the free boson theory in which we change the boundary condition from
Neumann to Dirichlet. Besides the bootstrap approach, direct infinite
and finite volume field theoretical calculations are presented, and
the relation to the open-closed string vertex \cite{Lucietti:2003ki,Lucietti:2004wy}
is demonstrated.

\section{Form factor axioms for boundary changing operators\label{sec:Form_factor_axioms}}

In this section we formulate the axioms, which have to be satisfied
by the matrix elements of local boundary condition changing operators.
We start by describing an integrable boundary system with a given
boundary condition, and focus on changing of the boundary condition
afterward. The calculation of two point function is also considered.

\subsection{Integrable boundary systems }

The Hilbert space of an integrable boundary system consists of multi-particle
states labeled by the particles' rapidities and their particle types.
For simplicity we analyze theories containing only one particle type
with a given mass $m$. Particles are then characterized only by their
rapidities, such that their energy and momentum are 
\begin{equation}
E=m\cosh\theta\quad,\qquad p=m\sinh\theta.
\end{equation}
Asymptotic \emph{in} states are prepared in the remote past, when
particles get far away form each other and from the boundary, which
we put on the right of the half-space at $x=0$. This well separated
particle state is equivalent to a free multi-particle state, which
we denote by 
\begin{equation}
\vert\theta_{1},\theta_{2},\dots,\theta_{n}\rangle_{in}^{\alpha}\quad,\qquad\theta_{1}>\theta_{2}>\dots>\theta_{n}>0
\end{equation}
where $\alpha$ labels the boundary condition. 

For $t\to+\infty$ all scatterings and reflections are terminated,
the particles are again far away from each other and from the boundary
forming the \emph{out} state, 

\begin{equation}
\vert\theta'_{1},\theta'_{2},\dots,\theta'_{m}\rangle_{out}^{\alpha}\quad,\qquad\theta'_{1}<\theta'_{2}<\dots<\theta'_{m}<0
\end{equation}
which is again equivalent to a free state. The two sets of states
form a complete basis separately and are connected by the multiparticle
reflection matrix. In an integrable theory, this reflection matrix
factorizes into the product of pairwise bulk scatterings and individual
reflections 
\begin{equation}
\vert\theta_{1},\theta_{2},\dots,\theta_{n}\rangle_{in}^{\alpha}=\prod_{i<j}S(\theta_{i}-\theta_{j})S(\theta_{i}+\theta_{j})\prod_{i}R^{\alpha}(\theta_{i})\vert-\theta_{1},-\theta_{2},\dots,-\theta_{n}\rangle_{out}^{\alpha}
\end{equation}
where $S(\theta_{i}-\theta_{j})$ connects the two particle asymptotic
\emph{in} and \emph{out} states in the bulk theory

\begin{tabular}{cc}
\begin{tabular}{c}
$\vert\theta_{1},\theta_{2}\rangle_{in}^{bulk}=S(\theta_{1}-\theta_{2})\vert\theta_{2},\theta_{1}\rangle_{out}^{bulk}\qquad\qquad$depicted
as\tabularnewline
\tabularnewline
\end{tabular} & %
\begin{tabular}{c}
\tabularnewline
\includegraphics[height=2cm]{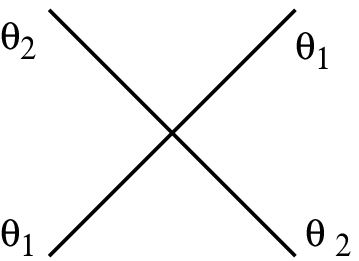}\tabularnewline
\end{tabular}\tabularnewline
\end{tabular}

It is defined originally for $\theta_{1}>\theta_{2}$ but can be analytically
continued for complex rapidity parameters such that the extended function
(denoted the same way) is meromorphic and satisfies unitarity and
crossing symmetry 
\begin{equation}
S(\theta)S(-\theta)=1\quad,\qquad S(i\pi-\theta)=S(\theta)
\end{equation}
It might have poles on the imaginary axis at locations $\theta=iu_{j}$
with residue $-i\textrm{res}_{\theta=iu_{j}}S(\theta)=\Gamma_{j}^{2}$,
some of which correspond to bound states. 

The amplitude $R^{\alpha}(\theta)$ connects the one particle asymptotic
states in the boundary theory 

\begin{tabular}{cc}
\begin{tabular}{c}
$\vert\theta\rangle_{in}^{\alpha}=R^{\alpha}(\theta)\vert-\theta\rangle_{out}^{\alpha}\qquad\qquad$depicted
as \tabularnewline
\tabularnewline
\end{tabular} & %
\begin{tabular}{c}
\tabularnewline
\hspace{2cm}\includegraphics[height=3cm]{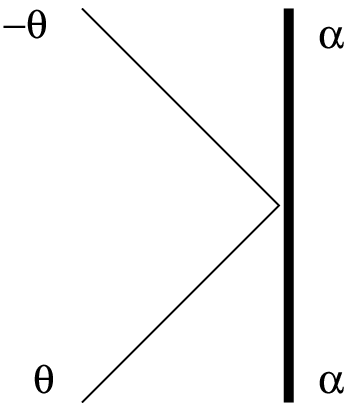}\tabularnewline
\end{tabular}\tabularnewline
\end{tabular}

It can also be extended from the fundamental domain $\theta>0$ to
a meromorphic function on the whole complex $\theta$ plane satisfying
unitarity and boundary crossing unitarity 
\begin{equation}
R^{\alpha}(\theta)R^{\alpha}(-\theta)=1\quad,\qquad R^{\alpha}(i\pi-\theta)S(2\theta)=R^{\alpha}(\theta)
\end{equation}
$R^{\alpha}(\theta)$ may have poles at imaginary locations $\theta=iv_{j}$
($0<v_{j}<\pi/2$), with residues $i\tilde{g}^{2}/2$, some corresponding
to excited boundary states. If the interpolating field has a nontrivial
vacuum expectation value then generally there is also a pole at $\theta=i\pi/2$
with residue 
\begin{equation}
-i\mathop{\textrm{Res}}_{\theta=\frac{i\pi}{2}}R^{\alpha}(\theta)=\frac{g_{\alpha}^{2}}{2}.\label{eq:defg}
\end{equation}

\subsection{Boundary changing operators}

A boundary condition changing operator, $\mathcal{O}_{\beta\alpha}(t)$
is a local operator, inserted at $t$, which changes the boundary
condition from $\alpha$, valid for time smaller than $t$, to $\beta$,
valid for times large than $t$ . Graphically it is represented as

\begin{center}
\includegraphics[height=3cm]{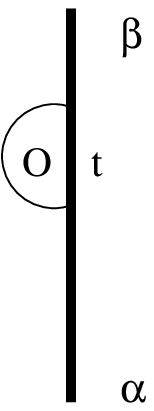}
\par\end{center}

The form factor of this boundary condition changing operator is defined
by its matrix element between asymptotic states related to the boundary
conditions $\alpha$ and $\beta$. We expect that the Hamiltonians
valid before and after the insertion can be used to transport the
operator in time, such that 
\begin{eqnarray}
\,_{out}^{\beta}\langle\theta'_{m},\dots,\theta'_{2},\theta'_{1}\vert\mathcal{O}_{\beta\alpha}(t)\vert\theta_{1},\theta_{2},\dots,\theta_{n}\rangle_{in}^{\alpha} & =\\
 &  & \negthickspace\negthickspace\negthickspace\negthickspace\negthickspace\negthickspace\negthickspace\negthickspace\negthickspace\negthickspace\negthickspace\negthickspace\negthickspace\negthickspace\negthickspace\negthickspace\negthickspace\negthickspace\negthickspace\negthickspace\negthickspace\negthickspace\negthickspace\negthickspace F_{mn}^{\mathcal{O}_{\beta\alpha}}(\theta'_{m},\dots,\theta'_{1};\theta_{1},\dots,\theta_{n})e^{-it(m\sum\cosh\theta_{i}+\Delta E_{{\rm bdry}}^{\beta\alpha}-m\sum\cosh\theta'_{j})}\nonumber 
\end{eqnarray}
where the difference in the boundary energies is $\Delta E_{{\rm bdry}}^{\beta\alpha}=E_{\alpha}-E_{\beta}$.
From now on we focus on the $t$-independent form factor $F_{mn}^{\mathcal{O}_{\beta\alpha}}$.
It is defined originally for $\theta_{1}>\theta_{2}>\dots>\theta_{n}>0$
and $\theta'_{1}<\theta'_{2}<\dots<\theta'_{m}<0$, but can be analytically
continued for any orderings and signs of the rapidities, and also
for complex values.

In \cite{Bajnok:2006ze} the form factors of a local boundary operator
were related to the correlation functions of the boundary theory via
the boundary reduction formula. The idea of the reduction formula
is that for large negative time the finite energy configurations contain
well localized separated particle states being far from each other
and from the boundary, thus forming an excitation of the free theory.
The interaction in this limit can be switched off adiabatically and
the interacting quantum field agrees with the free field up to the
wave-function renormalization constant. The particle creation operator,
expressed in terms of the free field, can be traded for the interpolating
field and the locality of the operator insertion guaranties a domain
of convergence for the continuation of the form factor in the complex
rapidity plane. Applying the same procedure for an outgoing state
and comparing the two expressions a crossing relation can be obtained
between the two form factors. By replacing the local boundary operator
with a local boundary \emph{changing} operator the continuity of the
interpolating field is not changed and similar argumentations can
be applied, which leads to the crossing formula 
\begin{equation}
F_{mn}^{\mathcal{O}_{\beta\alpha}}(\theta'_{m},\dots,\theta'_{2},\theta'_{1};\theta_{1},\theta_{2},\dots,\theta_{n})=F_{m-1n+1}^{\mathcal{O}_{\beta\alpha}}(\theta'_{m},\dots,\theta'_{2};\theta'_{1}+i\pi,\theta_{1},\theta_{2},\dots,\theta_{n})+\textrm{disc.}\label{eq:crossing}
\end{equation}
where disc. represents disconnected terms appearing whenever $\theta'_{1}$
equals any of the $\theta_{i}$. As a result of this crossing transformation
we can express all form factors in terms of the elementary form factors
\begin{equation}
\,_{out}^{\beta}\langle0\vert\mathcal{O}(0)\vert\theta_{1},\theta_{2},\dots,\theta_{n}\rangle_{in}^{\alpha}=F_{n}^{\mathcal{O}_{\beta\alpha}}(\theta_{1},\theta_{2},\dots,\theta_{n})
\end{equation}
Let us note that boundary form factors $F_{n}^{\mathcal{O}_{\beta\alpha}}(\theta_{1},\dots,\theta_{n})$
do depend in general on all the rapidities $\theta_{i}$, not just
on their differences, as the boundary breaks the Lorentz invariance.

\subsection{Axioms for the elementary form factors}

The form factor properties can be formally derived from the Zamolodchikov-Faddeev
algebra, see Appendix \ref{sec:ZF_algebra} and also\cite{Lesage:1998hh}.
We take these properties as axioms, such that functions satisfying
them determine local boundary changing operators completely. 

I. Permutation:
\begin{equation}
F_{n}^{\mathcal{O}_{\beta\alpha}}(\theta_{1},\dots,\theta_{i},\theta_{i+1},\dots,\theta_{n})=S(\theta_{i}-\theta_{i+1})F_{n}^{\mathcal{\mathcal{O}_{\beta\alpha}}}(\theta_{1},\dots,\theta_{i+1},\theta_{i},\dots,\theta_{n})
\end{equation}

\begin{center}
\includegraphics[height=3cm]{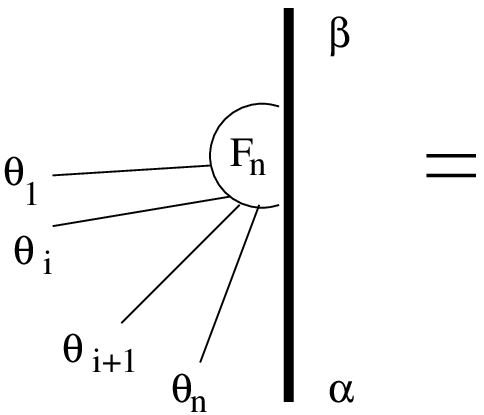}~~~\includegraphics[height=3cm]{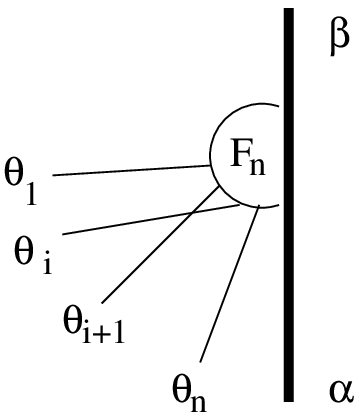}
\par\end{center}

II. Reflection:
\begin{equation}
F_{n}^{\mathcal{\mathcal{O}_{\beta\alpha}}}(\theta_{1},\dots,\theta_{n-1},\theta_{n})=R^{\alpha}(\theta_{n})F_{n}^{\mathcal{\mathcal{O}_{\beta\alpha}}}(\theta_{1},\dots,\theta_{n-1},-\theta_{n})
\end{equation}

\begin{center}
\includegraphics[height=3cm]{bd1}~~~ \includegraphics[height=33mm]{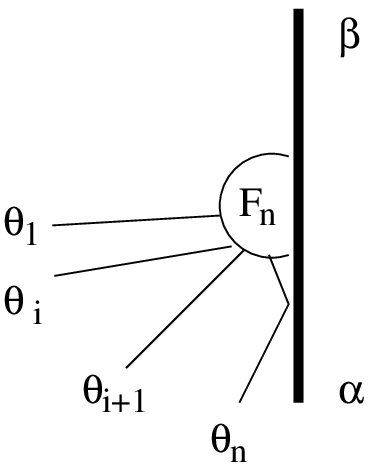}
\par\end{center}

III. Crossing reflection: 
\begin{equation}
F_{n}^{\mathcal{O}_{\beta\alpha}}(\theta_{1},\theta_{2},\dots,\theta_{n})=R^{\beta}(i\pi-\theta_{1})F_{n}^{\mathcal{O}_{\beta\alpha}}(2i\pi-\theta_{1},\theta_{2},\dots,\theta_{n})
\end{equation}

\begin{center}
\includegraphics[height=3cm]{bd1}~~~\includegraphics[height=3cm]{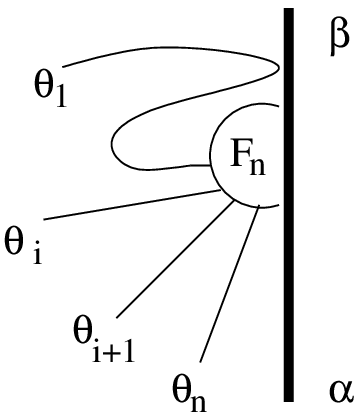}
\par\end{center}

The singularity structure of the form factors is determined on physical
grounds and can be axiomatized as follows:

IV. Kinematical singularity: 
\begin{equation}
-i\mathop{\textrm{Res}}_{\theta=\theta'}F_{n+2}^{\mathcal{\mathcal{O}_{\beta\alpha}}}(-\theta+i\pi,\theta',\theta_{1},\dots,\theta_{n})=\Big(R^{\beta}(\theta)-\prod_{i=1}^{n}S(\theta-\theta_{i})R^{\alpha}(\theta)S(\theta+\theta_{i})\Big)F_{n}^{\mathcal{\mathcal{O}_{\beta\alpha}}}(\theta_{1},\dots,\theta_{n})
\end{equation}

\begin{center}
\includegraphics[height=3cm]{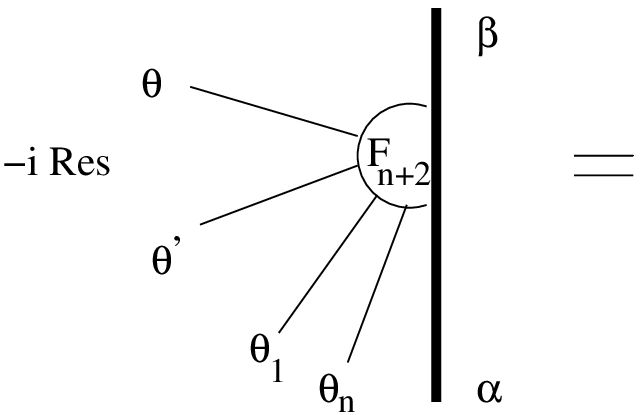}~~~\includegraphics[height=3cm]{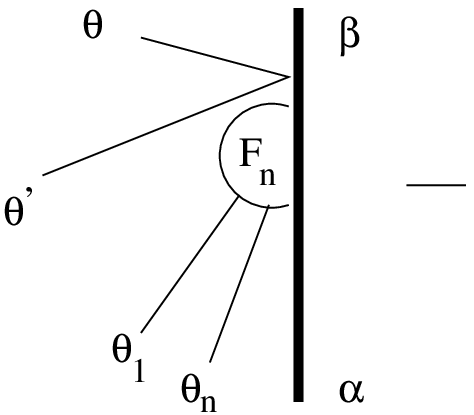}\includegraphics[height=3cm]{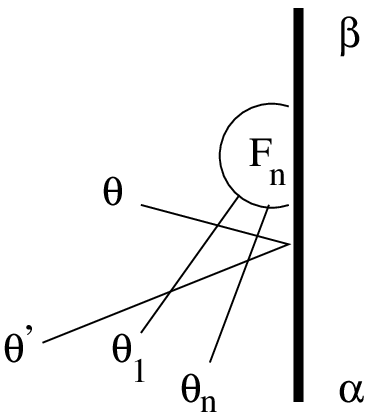}
\par\end{center}

V. Boundary kinematical singularity:

\begin{equation}
-i\mathop{\textrm{Res}}_{\theta=0}F_{n+1}^{\mathcal{O}_{\beta\alpha}}(\theta+\frac{i\pi}{2},\theta_{1},\dots,\theta_{n})=\Bigl(\frac{g_{\beta}}{2}-\frac{g_{\alpha}}{2}\prod_{i=1}^{n}S\bigl(\frac{i\pi}{2}-\theta_{i}\bigr)\Bigr)F_{n}^{\mathcal{O}_{\beta\alpha}}(\theta_{1},\dots,\theta_{n})
\end{equation}

\begin{center}
\includegraphics[height=3cm]{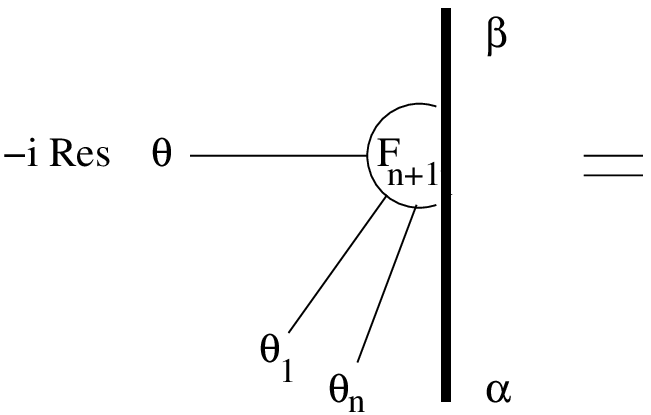}~~~\includegraphics[height=3cm]{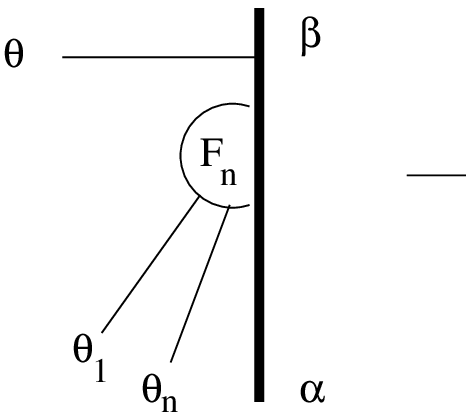}\includegraphics[height=3cm]{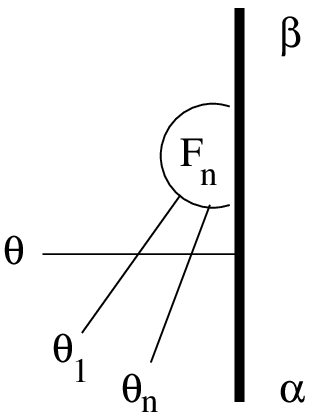}
\par\end{center}

VI. Bulk dynamical singularity: 
\begin{equation}
-i\mathop{\textrm{Res}}_{\theta=\theta'}F_{n+2}^{\mathcal{O}_{\beta\alpha}}(\theta+iu,\theta'-iu,\theta_{1},\dots,\theta_{n})=\Gamma F_{n+1}^{\mathcal{O}_{\beta\alpha}}(\theta,\theta_{1},\dots,\theta_{n})
\end{equation}

\begin{center}
\includegraphics[height=3cm]{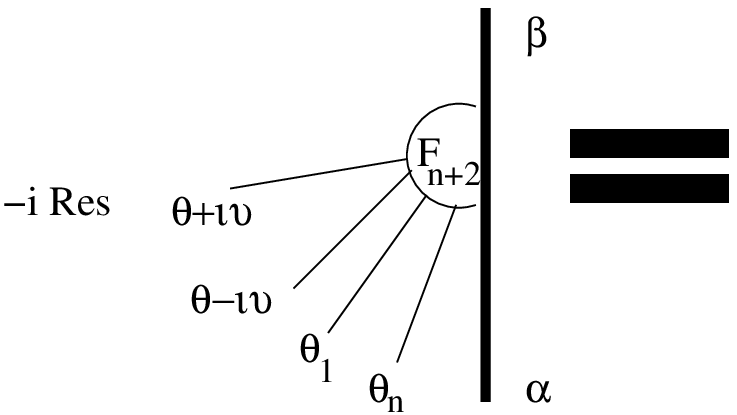}~~~\includegraphics[height=3cm]{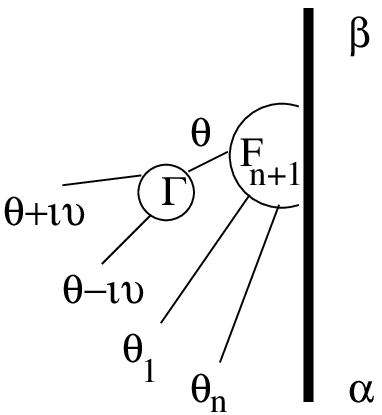}
\par\end{center}

VII. Boundary dynamical singularity:

\begin{equation}
-i\mathop{\textrm{Res}}_{\theta=iv}F_{n+1}^{\mathcal{O}_{\beta\alpha}}(\theta_{1},\dots,\theta_{n},\theta)=\tilde{g}_{\alpha}\tilde{F}^{\mathcal{\mathcal{O}_{\beta\alpha}}}(\theta_{1},\dots,\theta_{n}).
\end{equation}

\begin{center}
\includegraphics[height=3cm]{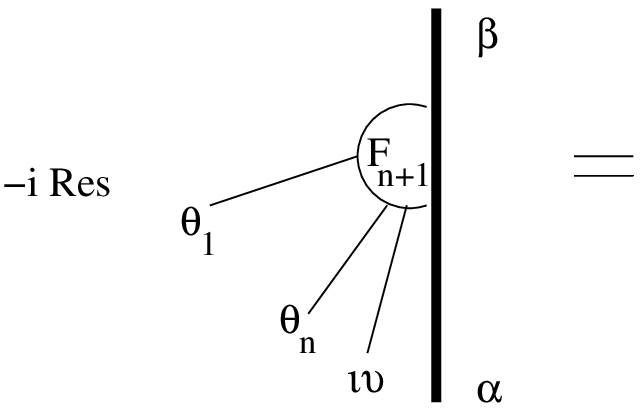}~~~\includegraphics[height=3cm]{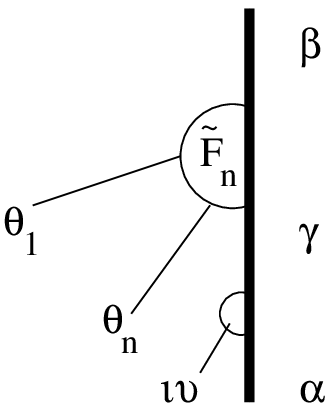} 
\par\end{center}

We would like to remark here that the axioms, except the boundary
kinematical singularity axiom, has already been proposed in \cite{Lesage:1998hh},
derived from the Zamolodchikov-Faddeev algebra, in a similar fashion
as presented in Appendix \ref{sec:ZF_algebra}. The boundary kinematical
singularity axiom is crucial as it can differentiate between physically
different boundary condition having the same reflection factor but
different sign of $g$.

\subsection{Consistency checks}

First we note that these axioms reduce to the form factor axioms of
local boundary operators in the $\alpha=\beta$ case. Furthermore,
we can also perform the same consistency checks, which were done for
the boundary form factors in \cite{Bajnok:2006ze}. Let us note that
the axioms are self-consistent in the sense that for specific rapidities
the $n+2$ particle form factor can be connected to the $n$ particle
form factor either by the kinematical singularity equations or by
using twice the boundary kinematical equations, and the two procedures
give the same result. Indeed taking double residue in the first case,
first at $\theta=\theta'$ and then at $\theta=i\frac{\pi}{2}$ gives
\begin{eqnarray}
i\mathop{\textrm{Res}}_{\theta=\frac{i\pi}{2}}i\mathop{\textrm{Res}}_{\theta'=\theta}F_{n+2}^{\mathcal{\mathcal{O}_{\beta\alpha}}}(-\theta+i\pi,\theta',\theta_{1},\dots,\theta_{n}) & =\\
 &  & \hspace{-5cm}=\left(-i\mathop{\textrm{Res}}_{\theta=\frac{i\pi}{2}}\right)\left(R^{\beta}(\theta)-R^{\alpha}(\theta)\prod_{i=1}^{n}S(\frac{i\pi}{2}-\theta_{i})S(\frac{i\pi}{2}+\theta_{i})\right)F_{n}^{\mathcal{\mathcal{O}_{\beta\alpha}}}(\theta_{1},\dots,\theta_{n}).\nonumber 
\end{eqnarray}
Taking now the residue at $\theta=\frac{i\pi}{2}$ first then at $\theta'=\frac{i\pi}{2}$
and using that $S(0)=-1$ gives 
\begin{eqnarray}
i\mathop{\textrm{Res}}_{\theta=\frac{i\pi}{2}}i\mathop{\textrm{Res}}_{\theta'=\frac{i\pi}{2}}F_{n+2}^{\mathcal{\mathcal{O}_{\beta\alpha}}}(-\theta+i\pi,\theta',\theta_{1},\dots,\theta_{n}) & =\\
 &  & \hspace{-5cm}=\left(\frac{g_{\beta}}{2}+\frac{g_{\alpha}}{2}\prod_{i=1}^{n}S\bigl(\frac{i\pi}{2}-\theta_{i}\bigr)\right)\left(\frac{g_{\beta}}{2}-\frac{g_{\alpha}}{2}\prod_{i=1}^{n}S\bigl(\frac{i\pi}{2}-\theta_{i}\bigr)\right)F_{n}^{\mathcal{\mathcal{O}_{\beta\alpha}}}(\theta_{1},\dots,\theta_{n}).\nonumber 
\end{eqnarray}
Combining the crossing symmetry of the S-matrix with the definition
of $g$ (\ref{eq:defg}) the two expressions are easily seen to be
equivalent.

There is another consistency check of the axioms, if one of the boundary
conditions can be obtained from the other by binding a particle to
it. This does not necessarily mean a boundary bound-state form factor,
as many boundary conditions can be obtained by placing an integrable
defect in front of a boundary \cite{Bajnok:2007jg}. If $R^{\beta}$
denotes the reflection factor of an integrable boundary condition
and $T_{\pm}(\theta)$ the left/right transmission factor of an integrable
defect then the reflection factor of the dressed boundary is
\begin{equation}
R^{\alpha}(\theta)=T_{-}(\theta)R^{\beta}(\theta)T_{+}(\theta)\label{eq:Rdress}
\end{equation}
A particle with imaginary rapidity, $\theta_{0}$, can always be considered
as an integrable defect $T_{\mp}(\theta)=S(\theta\mp\theta_{0})$
and in this case the dressed boundary reflection factor is 
\begin{equation}
R^{\alpha}(\theta)=S(\theta-\theta_{0})S(\theta+\theta_{0})R^{\beta}(\theta)\label{eq:RSS}
\end{equation}
which formally looks like a boundary excited reflection factor. One
example for this situation is the scaling Lee-Yang model with integrable
boundary conditions. There are two types of boundary conditions: the
$\mathbb{\beta=I}$ identity boundary condition, which does not allow
any bound-state and the $\alpha=\Phi$ boundary condition, which carries
a label $b$, and can be realized in the above sense with $\theta_{0}=\frac{i\pi(3-b)}{6}$.
The implementation of binding a particle with rapidity $\theta_{0}$
to the $\beta$ boundary is to consider the form factor equations
for $F_{n+1}^{\mathcal{O}_{\beta\beta}}(\theta_{1},\dots,\theta_{n},\theta_{0})$
in the rapidities $\theta_{1},\dots,\theta_{n}$ only. We claim that
the equations are the same as we presented above for $F_{n}^{\mathcal{O}_{\beta\alpha}}(\theta_{1},\dots,\theta_{n})$.
The permutation and crossing reflection axioms are trivially the same.
For the reflection axiom we move $\theta_{n}$ through $\theta_{0}$,
use the reflection axiom of the $\beta=\alpha$ case and move back
the reflected $-\theta_{n}$ through $\theta_{0}$. As a result we
obtain the dressed reflection factor (\ref{eq:RSS}). The singularity
axioms can easily be seen to be the same, too. Let us note that although
all the equations for $F_{n}^{\mathcal{O}_{\beta\alpha}}(\theta_{1},\dots,\theta_{n})$
appear as equations for $F_{n+1}^{\mathcal{O}_{\beta\beta}}(\theta_{1},\dots,\theta_{n},\theta_{0})$,
the latter one satisfies additional axioms, such as the permutation
or reflection axiom involving $\theta_{0}$, thus we do not expect
the two form factors to be equal.

\subsection{General solution of the form factor axioms}

We start this section by determining the one particle form factor
and use later this solution to construct the general multiparticle
form factor. In order to simplify notations we suppress the operator
$\mathcal{O}_{\beta\alpha}$ in the index of the form factor and write
only $\beta\alpha$ explicitly. 

The equations for the one particle form factor read%
\footnote{These equations had been also found in \cite{Lesage:1998hh} in the
context of the sinh-Gordon theory%
}: 
\begin{equation}
F_{1}^{\beta\alpha}(\theta)=R^{\alpha}(\theta)F_{1}^{\beta\alpha}(-\theta)\quad;\quad F_{1}^{\beta\alpha}(i\pi+\theta)=R^{\beta}(-\theta)F_{1}^{\beta\alpha}(i\pi-\theta),\label{eq:1pff}
\end{equation}
where the reflection amplitudes $R^{\alpha}(\theta)$, $R^{\beta}(\theta)$
are assumed to be meromorphic. From general considerations we assume
that $F_{1}^{\beta\alpha}(\theta)$ is analytic on $0\leq\Im{\rm m}(\theta)\leq\pi$.
The construction of solving (\ref{eq:1pff}) is reduced to a problem
already solved in the bulk form factor bootstrap. To this end we write
\begin{equation}
F_{1}^{\beta\alpha}(\theta)=h^{\alpha}(\theta)h^{\beta}(i\pi-\theta)
\end{equation}
and suppose that 
\begin{equation}
h^{\gamma}(\theta)=R^{\gamma}(\theta)h^{\gamma}(-\theta)\quad,\qquad h^{\gamma}(i\pi+\theta)=h^{\gamma}(i\pi-\theta)\quad,\qquad\gamma=\alpha,\beta\label{eq:g1}
\end{equation}
which are nothing else but the bulk two particle form factor equations
\cite{Karowski:1978vz}, where the reflection amplitude, $R^{\gamma}(\theta)$,
plays the role of the S-matrix. To obtain a solution of (\ref{eq:g1})
we use the theorem of Karowski and Weisz \cite{Karowski:1978vz}.
Assume that the function $h(\theta)$ is meromorphic in the physical
strip $0\leq\Im{\rm m}(\theta)<\pi$ with possible poles at $i\alpha_{1},\dots,i\alpha_{l}$
and zeros at $i\beta_{1},\dots,i\beta_{k}$ and grows as at most a
polynomial in $\exp(|\theta|)$ for $|\Re{\rm e}\,\theta|\rightarrow\infty$,
furthermore it satisfies 
\begin{equation}
h(\theta)=R(\theta)h(-\theta)\ ,\quad h(i\pi-\theta)=h(i\pi+\theta)\ ,\quad R(\theta)=\exp\left\{ \int_{0}^{\infty}dt\, f(t)\sinh\left(\frac{t\theta}{i\pi}\right)\right\} 
\end{equation}
then it is uniquely defined up to normalization as 
\begin{equation}
h(\theta)=\frac{\prod_{j=1}^{k}\sinh\left(\frac{1}{2}(\theta-i\beta_{j})\right)\sinh\left(\frac{1}{2}(\theta+i\beta_{j})\right)}{\prod_{j=1}^{l}\sinh\left(\frac{1}{2}(\theta-i\alpha_{j})\right)\sinh\left(\frac{1}{2}(\theta+i\alpha_{j})\right)}\exp\left\{ \int_{0}^{\infty}dt\, f(t)\frac{\sin^{2}\left(\frac{i\pi-\theta}{2\pi}t\right)}{\sinh t}\right\} .
\end{equation}
In the typical applications the reflection amplitude can be expressed
as products of the blocks, $(x_{i})$, 
\begin{equation}
R^{\gamma}(\theta)=\prod_{i=1}^{k}(x_{i}^{\gamma})\ ,\ -(x)=-\frac{\sinh(\frac{\theta}{2}+i\frac{\pi x}{2})}{\sinh(\frac{\theta}{2}-i\frac{\pi x}{2})}=\exp\left\{ 2\int_{0}^{\infty}\frac{dt}{t}\frac{\sinh t(1-x)}{\sinh t}\sinh\left(\frac{t\theta}{i\pi}\right)\right\} \label{eq:blockdef}
\end{equation}
where $0<x<1$. The validity of this integral representation can be
extended by periodicity $(x\pm2)=(x)$ and by the relation $(-x)=(x)^{-1}$.
Thus the minimal solution, corresponding to $(-1)^{k}R^{\gamma}(\theta)$
is given as
\begin{equation}
h^{\gamma}(\theta)=\exp\left\{ 2\int_{0}^{\infty}\frac{dt}{t}\frac{\sum_{i=1}^{k}\sinh\left(t(1-x_{i}^{\gamma})\right)}{\sinh^{2}t}\sin^{2}\left(\frac{i\pi-\theta}{2\pi}t\right)\right\} 
\end{equation}
if $k$ is even. In case of odd $k$, due to the extra minus sign
in $R^{\gamma}(\theta)$, the minimal solution $h^{\gamma}(\theta)$
necessarily contains a zero at the origin which can be implemented
by putting an extra $\sinh\frac{\theta}{2}$ into it.

We would like to remark that the one-particle minimal boundary changing
form factors (\ref{eq:1pff}) have been found, by slightly different
methods, in the off-critical Ising model, the sinh-Gordon model and
for double well problem of dissipative quantum mechanics \cite{Lesage:1997tc,Lesage:1998hh}.

Note that if $F_{1}^{\beta\alpha}(\theta)$ is a solution of (\ref{eq:1pff})
then $F_{1}^{\beta\alpha}(\theta)Q(\theta)$ is also a solution provided
$Q(\theta)=Q(-\theta)$ and $Q(i\pi+\theta)=Q(i\pi-\theta)$, i.e.
if $Q$ is even and $2\pi i$ periodic. Therefore one can assume that
$Q$ is the function of $y=e^{\theta}+e^{-\theta}$. Thus the general
solution of eq. (\ref{eq:1pff}) can be written as 
\begin{equation}
F_{1}^{\beta\alpha}(\theta)=r^{\beta\alpha}(\theta)Q_{1}(y),\qquad y=e^{\theta}+e^{-\theta},
\end{equation}

The general form of the multi-particle form factors which, additionally
to the reflection equations, satisfies also the permutation and the
singularity equations, can be written in the following form%
\footnote{This parametrization was found also in \cite{Lesage:1998hh} for the
off-critical Ising and sinh-Gordon model.%
}: 
\begin{equation}
F_{n}^{\beta\alpha}(\theta_{1},\theta_{2},\dots,\theta_{n})=H_{n}\prod_{i=1}^{n}\frac{r^{\beta\alpha}(\theta_{i})}{y_{i}}\prod_{i<j}\frac{f(\theta_{i}-\theta_{j})f(\theta_{i}+\theta_{j})}{(y_{i}+y_{j})}Q_{n}(y_{1},y_{2}\dots,y_{n}).\label{eq:GenAnsatz}
\end{equation}
Here $f(\theta)$ is the minimal bulk two particle form factor, defined
as the minimal solution, i.e. the one with the least poles and zeros
compatible with the dynamics of the theory, of the equations
\begin{equation}
f(\theta)=S(\theta)f(-\theta)\quad,\qquad f(i\pi-\theta)=f(i\pi+\theta).
\end{equation}
As a consequence of the form factor equations, $Q_{n}$ is a $2\pi i$
periodic, symmetric and even function of the rapidities, $\theta_{i}$,
i.e. it is symmetric in the variable $y_{i}=2\cosh\theta_{i}$. The
denominator $\prod_{i}y_{i}$ is responsible for the boundary, while
the product $\prod_{i<j}(y_{i}+y_{j})$ for the bulk kinematical singularity.
The boundary and bulk kinematical singularity axioms result in recursions
relating $Q_{n}$ to $Q_{n-1}$ and $Q_{n-2}$, respectively. The
bulk dynamical pole equation relates also $Q_{n}$ to $Q_{n-1}$ if
it is present. The corresponding pole is usually included in $f(\theta)$.

An important restriction on the form factor functions follows from
requiring a power law bounded ultraviolet behaviour for the two point
correlator of two boundary changing operators $\langle0|\mathcal{O}^{\gamma\beta}(\tau)\mathcal{O}^{\beta\alpha}(0)|0\rangle$:
the growth of the function $F_{n}^{\beta\alpha}(\theta_{1},\dots,\theta_{n})$
must be bounded by some exponential of the rapidity as $\theta\rightarrow\infty$
(i.e. the form factors only grow polynomially with particle energy).
If $r\left(\theta\right)$ and $f\left(\theta\right)$ are specified
in a way to include all poles induced by the dynamics of the model,
then it follows that the functions $Q_{n}$ must be \textsl{polynomials}
of the variables $y_{i}$.

\subsection{Two-point function}

Once an appropriate solution of the form factor axioms is found, it
can be used to describe correlators of boundary changing operators.
The two-point function of boundary changing operators can be computed
by inserting a complete set of states 
\begin{equation}
\langle0\vert\mathcal{O}^{\gamma\beta}(t)\mathcal{O}^{\beta\alpha}(0)\vert0\rangle=\sum_{n=0}^{\infty}\frac{1}{(2\pi)^{n}}\int_{\theta_{1}>\theta_{2}>\dots>\theta_{n}>0}d\theta_{1}d\theta_{2}\dots d\theta_{n}e^{-it\Delta E_{{\rm bdry}}^{\gamma\beta}-imt\sum_{i}\cosh\theta_{i}}F_{n}^{\gamma\beta}F_{n}^{\beta\alpha+}\label{eq:2pt}
\end{equation}
where time translation covariance was used, and the form factors were
abbreviated by 
\begin{equation}
F_{n}^{\beta\alpha}=\langle0\vert\mathcal{O}^{\beta\alpha}(0)\vert\theta_{1},\theta_{2},\dots,\theta_{n}\rangle_{in}=F_{n}^{\beta\alpha}(\theta_{1},\theta_{2},\dots,\theta_{n})
\end{equation}
and by 
\begin{equation}
F_{n}^{\beta\alpha+}=\,_{in}\langle\theta_{1},\theta_{2},\dots,\theta_{n}\vert\mathcal{O}^{\beta\alpha}(0)\vert0\rangle=F_{n}^{\beta\alpha}(i\pi+\theta_{n},i\pi+\theta_{n-1},\dots,i\pi+\theta_{1}).
\end{equation}
The latter one, for unitary theories, is the complex conjugate of
the first one: $F_{n}^{+}=F_{n}^{*}$. In the Euclidean $(r=it)$
version of the theories the form factor expansion of the correlator
for large separations converges rapidly since multi-particle terms
are exponentially suppressed.

\section{Model studies\label{sec:Model_studies}}

In this section we explicitly carry out the form factor bootstrap
program in the free boson and Lee-Yang models.

\subsection{Free boson with linear boundary conditions\label{sub:Free_boson}}

As a first step we carry out the form factor bootstrap program and
calculate explicitly the form factors of the operators, which change
the linear boundary condition with parameter $\lambda^{\alpha}$ to
that of with $\lambda^{\beta}$. When the boundary is changed from
Neumann to Dirichlet we recover the same result from the direct solution
of the model.

\subsubsection{Solution of the form factor equation}

The reflection factor of the free boson with linear boundary condition
has the form 
\begin{equation}
R^{\gamma}(\theta)=\frac{\sinh\theta-i\lambda^{\gamma}}{\sinh\theta+i\lambda^{\gamma}}
\end{equation}
Following the general strategy, we search for the one particle form
factor $F_{1}^{\beta\alpha}(\theta)$ in the form 
\begin{equation}
F_{1}^{\beta\alpha}(\theta)=r^{\beta\alpha}(\theta)Q_{1}(y)\quad,\qquad r^{\beta\alpha}(\theta)=h^{\alpha}(\theta)h^{\beta}(i\pi-\theta)\quad,\qquad y=e^{\theta}+e^{-\theta}
\end{equation}
where the functions $h^{\gamma}(\theta)$ satisfy 
\begin{equation}
h^{\gamma}(\theta)=R^{\gamma}(\theta)h^{\gamma}(-\theta)\quad,\qquad h^{\gamma}(i\pi+\theta)=h^{\gamma}(i\pi-\theta)\quad,\qquad\gamma=\alpha,\beta.
\end{equation}
As these equations are the same as the minimal two-particle form factor
equations in the sinh-Gordon theory, we borrow the results from there
\cite{Fring:1992pt}
\begin{equation}
h^{\gamma}(\theta)=\mathcal{N}^{\gamma}\exp\left\{ 4\int\frac{dt}{t}\frac{\sinh\left(tp^{\gamma}\right)\sinh\left(t(1-p^{\gamma})\right)}{\cosh(t)\,\sinh(2t)}\sin^{2}\left(\frac{t}{\pi}(i\pi-\theta)\right)\right\} 
\end{equation}
where $\lambda^{\gamma}=\sin\pi p^{\gamma}$. The normalization
\begin{equation}
\mathcal{N}^{\gamma}=\exp\left\{ -2\int\frac{dt}{t}\frac{\sinh\left(tp^{\gamma}\right)\sinh\left(t(1-p^{\gamma})\right)}{\cosh(t)\,\sinh(2t)}\right\} 
\end{equation}
is chosen such that the minimal form factor satisfies the following
identity
\begin{equation}
h^{\gamma}(\theta+i\pi)h^{\gamma}(\theta)=\frac{\sinh\theta}{\sinh\theta+i\lambda^{\gamma}}.
\end{equation}
Strictly speaking, this identification with the sinh-Gordon theory
is valid only if $p^{\gamma}\in[0,1]$, outside of this domain analytic
continuation is needed.

Since the scattering matrix is trivial, $S\equiv1$, and the reflection
factor does not have any pole at $\frac{i\pi}{2}$, the Ansatz for
the multiparticle form factor is 
\begin{equation}
F_{n}^{\beta\alpha}(\theta_{1},\theta_{2},\dots,\theta_{n})=\left\langle \mathcal{O}_{\beta\alpha}\right\rangle H_{n}Q_{n}(y_{1},\dots,y_{n})\prod_{i=1}^{n}r^{\beta\alpha}(\theta_{i})\prod_{i<j}\frac{1}{y_{i}+y_{j}}
\end{equation}
where $Q$ is a symmetric polynomial. When the reflection factors
are different the kinematical singularity axiom 
\begin{equation}
-i\mathop{\textrm{Res}}_{\theta=\theta'}F_{n+2}^{\beta\alpha}(-\theta+i\pi,\theta',\theta_{1},\dots,\theta_{n})=\left(R^{\beta}(\theta)-R^{\alpha}(\theta)\right)F_{n}^{\beta\alpha}(\theta_{1},\dots,\theta_{n})
\end{equation}
recursively links $Q_{n+2}$ to $Q_{n}$. Using that 
\begin{eqnarray}
r^{\beta\alpha}(-\theta+i\pi)r^{\beta\alpha}(\theta) & = & \frac{\sinh\theta}{\sinh\theta+i\lambda^{\alpha}}\frac{\sinh\theta}{\sinh\theta+i\lambda^{\beta}}\\
R^{\beta}(\theta)-R^{\alpha}(\theta) & = & \frac{2i\sinh\theta(\lambda^{\alpha}-\lambda^{\beta})}{(\sinh\theta+i\lambda^{\alpha})(\sinh\theta+i\lambda^{\beta})}
\end{eqnarray}
and choosing $H_{2n}=\left(4(\lambda^{\alpha}-\lambda^{\beta})\right)^{n}$
we obtain a recursion, connecting either the even or the odd particle
polynomials to each other, which reads as
\begin{equation}
Q_{n+2}(-y,y,y_{1},\dots,y_{n})=\prod_{i=1}^{n}(y+y_{i})(-y+y_{i})Q_{n}(y_{1},\dots,y_{n})
\end{equation}
Let us choose $Q_{0}=1$, and solve the first few equations explicitly

\begin{eqnarray}
Q_{2}(-y,y)=Q_{0} & \to & Q_{2}=1\\
Q_{4}(-y,y,y_{1},y_{2})=(y^{2}-y_{1}^{2})(y^{2}-y_{2}^{2})Q_{2}(y_{1},y_{2}) & \to & Q_{4}=\left((\sigma_{2}^{(4)})^{2}+\sigma_{1}^{(4)}\sigma_{3}^{(4)}-4\sigma_{4}^{(4)}\right)\nonumber 
\end{eqnarray}
where in the last line we introduced the elementary symmetric polynomials,
defined as
\begin{equation}
\prod_{i=1}^{n}(y+y_{i})=\sum_{k}y^{n-k}\sigma_{k}^{(n)}(y_{1},\dots,y_{n})\label{eq:symm_polys}
\end{equation}
With this definition we have $\sigma_{k}^{(n)}=0$ if $k<0$ or $k>n$.
In what follows we will usually omit the arguments of the symmetric
polynomials, if it does not lead to any confusion. It is instructive
to rewrite the solution by explicitly dividing by the product $\prod_{i<j}(y_{i}+y_{j})$:
\begin{equation}
G_{2}=\frac{Q_{2}}{y_{12}}=\frac{1}{y_{12}}\ ,\quad G_{4}=\frac{Q_{4}}{y_{12}y_{13}y_{14}y_{23}y_{24}y_{34}}=\frac{1}{y_{12}y_{34}}+\frac{1}{y_{13}y_{24}}+\frac{1}{y_{14}y_{23}}=\frac{1}{y_{34}}G_{2}+\mbox{perm.}
\end{equation}
 where $y_{ij}=y_{i}+y_{j}$. This solution generalizes to 
\begin{equation}
G_{n}=\frac{Q_{n}}{\prod_{i<j}y_{ij}}=\frac{1}{y_{nn-1}}G_{n-2}+\mbox{perm}=\sum_{\mathrm{all\, pairings}}\frac{1}{\prod_{\mathrm{pairs}(i,j)}y_{ij}}\label{eq:Boson_G}
\end{equation}
Strictly speaking (\ref{eq:Boson_G}) gives the solution for even
number of particles. However, similar calculation can be done for
the odd particle sector starting from $Q_{1}=1$, and finally one
arrives at the same formula (\ref{eq:Boson_G}), but in this case
a pairing means that one of the $y$'s is left unpaired and does not
contribute to the product. The resulting formula is very natural for
a free theory and reflects Wick theorem. Actually it is not hard to
see that this $G_{n}$ solves the recursion equations since in the
parametrization 
\begin{equation}
F_{n}^{\beta\alpha}(\theta_{1},\theta_{2},\dots,\theta_{n})=\left\langle \mathcal{O}_{\beta\alpha}\right\rangle H_{n}G_{n}(y_{1},\dots,y_{n})\prod_{i=1}^{n}r^{\beta\alpha}(\theta_{i})
\end{equation}
the kinematical recursion equation takes the form: 
\begin{equation}
\lim_{y_{n+2}\to-y_{n+1}}y_{n+1n+2}G_{n+2}(y_{1},\dots,y_{n},y_{n+1},y_{n+2})=G_{n}(y_{1},\dots,y_{n})
\end{equation}
which is satisfied by construction. In the following we try to directly
solve the same model.

\subsubsection{Direct solution of the model}

The free massive scalar field $\Phi(x,t)$ restricted to the negative
half-line $x\leq0$ subject to the linear boundary condition
\begin{equation}
\partial_{x}\Phi(x,t)|_{x=0}=-\lambda m\Phi(0,t)\label{eq:robinbc}
\end{equation}
can be described by the following Lagrangian:
\begin{equation}
\mathcal{L}=\Theta(-x)\left(\frac{1}{2}(\partial_{t}\Phi)^{2}-\frac{1}{2}(\partial_{x}\Phi)^{2}-\frac{m^{2}}{2}\Phi^{2}\right)-\delta(x)\frac{\lambda m}{2}\Phi^{2}
\end{equation}
This one parameter family of linear boundary conditions interpolates
between Neumann $\partial_{x}\Phi\vert_{x=0}=0$ (for $\lambda=0$)
and Dirichlet $\Phi\vert_{x=0}=0$ (for $\lambda\to\infty$) boundary
conditions and can be solved explicitly. The mode decomposition of
the field is 
\begin{equation}
\Phi(x,t)=\int_{0}^{\infty}\tilde{dk}\Bigl\{ a(k)e^{-i\omega(k)t}\phi_{k}(x)+a^{+}(k)e^{i\omega(k)t}\phi_{k}^{*}(x)\Bigr\}\ ;\quad\phi_{k}(x)=e^{ikx}+R(k)e^{-ikx}
\end{equation}
where $\tilde{dk}=\frac{dk}{4\pi\omega(k)}$ and creation/annihilation
operators are normalized as 
\begin{equation}
[a(k),a^{+}(k')]=4\pi\omega(k)\delta(k-k')\quad,\quad k,k'>0\label{eq:FB_algebra}
\end{equation}
with $\omega(k)=\sqrt{m^{2}+k^{2}}$, and the boundary condition fixes
the reflection factor to be 
\begin{equation}
R(k)=\frac{k-i\lambda m}{k+i\lambda m}\quad\longrightarrow\quad R(\theta)=\frac{\sinh\theta-i\lambda}{\sinh\theta+i\lambda}
\end{equation}
The vacuum is defined as 
\begin{equation}
a(k)\vert0\rangle=0\quad;\qquad k>0
\end{equation}
and the states are created by acting successively with the creation
operators $a^{+}(k)$. The wave functions are orthonormalized, satisfying
\begin{equation}
\int_{-\infty}^{0}\phi_{k}(x)\phi_{k'}^{*}(x)dx=2\pi\delta(k-k')\quad,\quad k,k'>0\quad;\qquad\phi_{k}^{*}(x)=R(-k)\phi_{k}(x)\label{eq:Orthogonality}
\end{equation}
and they also form a complete set
\begin{equation}
\int_{0}^{\infty}\frac{dk}{2\pi}\phi_{k}^{*}(x)\phi_{k}(y)=\delta(x-y)\label{eq:Completeness}
\end{equation}
These can be obtained by regularizing the integrals as 
\begin{equation}
\int_{-\infty}^{0}e^{ikx}dx=\lim_{\epsilon\to0}\int_{-\infty}^{0}e^{i(k-i\epsilon)x}dx=\lim_{\epsilon\to0}\frac{-i}{k-i\epsilon}=-i\mathbb{P}_{\frac{1}{k}}+\pi\delta(k)
\end{equation}

We now turn to the problem of changing the boundary condition. Let
us assume that for $t<0$ the boundary condition has label $\lambda^{\alpha},$
while for $t>0$ it is changed to $\lambda^{\beta}.$ The corresponding
reflection factors are denoted by $R^{\alpha}$ and $R^{\beta}$,
respectively. The expansion of the free field before and after the
insertion of the boundary changing operator is 
\begin{equation}
\Phi(x,t)=\begin{cases}
\int_{0}^{\infty}\tilde{dk}\Bigl\{ a_{\alpha}(k)e^{-i\omega(k)t}\phi_{k}^{\alpha}+a_{\alpha}^{+}(k)e^{i\omega(k)t}\phi_{k}^{\alpha*}\Bigr\} & \qquad t<0\\
\int_{0}^{\infty}\tilde{dk}\Bigl\{ a_{\beta}(k)e^{-i\omega(k)t}\phi_{k}^{\beta}+a_{\beta}^{+}(k)e^{i\omega(k)t}\phi_{k}^{\beta*}\Bigr\} & \qquad t>0
\end{cases}
\end{equation}

As each set of modes form a complete system, we can expand each in
terms of the other 
\begin{eqnarray}
A_{kk'}^{\alpha\beta} & \equiv & \int_{-\infty}^{0}\phi_{k}^{\alpha*}(x)\phi_{k'}^{\beta}(x)dx=\\
 & = & \frac{4k\, k'm(\lambda^{\alpha}-\lambda^{\beta})}{(k^{2}-k'^{2})(k-im\lambda^{\alpha})(k'+im\lambda^{\beta})}+\frac{2\pi(m^{2}\lambda^{\beta}\lambda^{\alpha}+k\, k')}{(k-im\lambda^{\alpha})(k'+im\lambda^{\beta})}\delta(k-k')\nonumber 
\end{eqnarray}
where the first term is understood in the principal value sense, and
$k,k'>0$. The creation/annihilation operators can be related by demanding
the continuity of the field $\Phi(x,t)$ and its momentum $\partial_{t}\Phi(x,t)=\Pi(x,t)$
at $t=0$: 
\begin{eqnarray}
\omega(k)\Phi(x,0)\pm i\Pi(x,0)=\nonumber \\
 &  & \hspace{-4cm}=\int_{0}^{\infty}\tilde{dk'}\left\{ a_{\gamma}(k')\left(\omega(k)\pm\omega(k')\right)+a_{\gamma}^{+}(k')R^{\gamma}(-k')\left(\omega(k)\mp\omega(k')\right)\right\} \phi_{k'}^{\gamma}(x)
\end{eqnarray}
where $\gamma$ can be either $\alpha$ or $\beta$. Comparing the
two expressions we can extract that
\begin{equation}
a_{\alpha}(k)=\int_{0}^{\infty}\tilde{dk'}\left\{ a_{\beta}(k')\left(\omega(k)+\omega(k')\right)+R^{\beta}(-k')a_{\beta}^{+}(k')\left(\omega(k)-\omega(k')\right)\right\} A_{kk'}^{\alpha\beta}\label{eq:a_alpha_with_beta}
\end{equation}
An important effect of the boundary changing operator is that it changes
the vacuum of the system: the vacuum for the $\alpha$ boundary condition,
$a_{\alpha}(k)\vert0\rangle^{\alpha}=0$, becomes a complicated excited
state for the $\beta$ boundary condition. As the transformation between
the modes is linear we face with a Boguliubov transformation, whose
solution has an exponential form
\begin{multline}
\vert0\rangle^{\alpha}=\mathcal{N}^{\alpha\beta}\left(1+\int_{0}^{\infty}\tilde{dk_{0}}K_{1}^{\alpha\beta}(k_{0})a_{\beta}^{+}(k_{0})\right)\times\hspace{4cm}\\
\times\exp\left\{ \frac{1}{2}\iint_{0}^{\infty}\tilde{dk_{1}}\tilde{dk_{2}}K_{2}^{\alpha\beta}(k_{1},k_{2})a_{\beta}^{+}(k_{1})a_{\beta}^{+}(k_{2})\right\} \vert0\rangle^{\beta}\label{eq:Vac_rel}
\end{multline}
where $K_{1}^{\alpha\beta}$ and $K_{2}^{\alpha\beta}$ are the solutions
of
\begin{equation}
\int_{0}^{\infty}\tilde{dk'}\left(\omega(k)+\omega(k')\right)A_{kk'}^{\alpha\beta}K_{1}^{\alpha\beta}(k')=0\label{eq:K1_condition}
\end{equation}
and

\begin{equation}
A_{kk'}^{\alpha\beta}R^{\beta}(-k')\left(\omega(k)-\omega(k')\right)+\int_{0}^{\infty}\tilde{dk_{1}}A_{kk_{1}}^{\alpha\beta}\left(\omega(k)+\omega(k_{1})\right)K_{2}^{\alpha\beta}(k_{1},k')=0.\label{eq:K2_condition}
\end{equation}
The normalization is the overlap of the two vacua $\mathcal{N}^{\alpha\beta}=\mathcal{N}^{\beta\alpha*}=\phantom{I}^{\beta}\langle0\vert0\rangle^{\alpha}$.
To see the validity of (\ref{eq:Vac_rel}) one may check first that
$a_{\alpha}(k)$ commutes with the factor in front of the exponential
provided (\ref{eq:K1_condition}) is satisfied. Then developing the
exponential into Taylor series and acting with the $\beta$-representation
of the $a_{\alpha}$ annihilation operator (\ref{eq:a_alpha_with_beta})
it is not hard to see order-by-order that it annihilates the state.
The equations (\ref{eq:K1_condition},\ref{eq:K2_condition}) seem
hard to solve, nevertheless one may check that the bootstrap solution
satisfies them.

Comparing the bosonic algebra (\ref{eq:FB_algebra}) to the free boson
Zamolodchikov-Faddeev algebra (\ref{eq:bulk_ZF}) shows that they
differ only in the normalization. We can thus relate the kernels $K_{1}^{\alpha\beta}$
and $K_{2}^{\alpha\beta}$ to the form factors, as
\begin{equation}
F_{1}^{\beta\alpha}(\theta)=\frac{1}{\sqrt{2}}\phantom{I}^{\beta}\langle0\vert a_{\alpha}^{+}(k)\vert0\rangle^{\alpha}=\frac{1}{\sqrt{2}}\mathcal{N}^{\alpha\beta}K_{1}^{\beta\alpha*}(k)
\end{equation}
and
\begin{equation}
F_{2}^{\beta\alpha}(\theta_{1},\theta_{2})=\frac{1}{2}\phantom{I}^{\beta}\langle0\vert a_{\alpha}^{+}(k_{1})a_{\alpha}^{+}(k_{2})\vert0\rangle^{\alpha}=\frac{1}{2}\mathcal{N}^{\alpha\beta}K_{2}^{\beta\alpha*}(k_{1},k_{2})
\end{equation}
with $k_{i}=m\sinh\theta_{i}$. Solving equations (\ref{eq:K1_condition},\ref{eq:K2_condition})
thus would also determine the form factors. However, solving these
equations is quite involved, we could not carry it out for the general
case. In Appendix \ref{sec:Neumann_to_Dirichlet} we considered the
case when we change the boundary condition from Neumann to Dirichlet.
By mapping the problem to the already solved open-closed string vertex
\cite{Lucietti:2003ki,Lucietti:2004wy}, we managed to read of the
solution which agrees with the bootstrap prediction.

\subsection{The boundary scaling Lee-Yang model}

The Lee-Yang theory is the simplest, non-unitary Conformal Field Theory,
the $\mathcal{M}_{2,5}$ minimal model, with the central charge $c=-\frac{22}{5}$.
The Virasoro algebra, $Vir$, has only two irreducible representation,
denoted by $V_{0}$ and $V_{h}$ with the highest weights $0$ and
$h=-\frac{1}{5}$. The periodic model carries the representation of
two copies of the Virasoro algebra, $Vir\otimes\overline{Vir}$ and
the modular invariance constrains the Hilbert space to decompose as
\begin{equation}
\mathcal{H}=V_{0}\otimes\overline{V}_{0}+V_{h}\otimes\overline{V}_{h}
\end{equation}
We denote the corresponding primary fields by $\mathbb{I}$ of the
scaling dimension $0$, and $\Phi$ of the scaling dimension $-\frac{2}{5}$,
respectively. 

A conformal boundary breaks the symmetry into a single Virasoro algebra.
We will denote the two conformal boundary conditions by $\mathbb{I}$-boundary
and $\Phi$-boundary. The corresponding Hilbert spaces decompose as
\begin{equation}
\mathcal{H}_{\mathbb{I}}=V_{0}\qquad,\qquad\mathcal{H}_{\Phi}=V_{0}+V_{h}
\end{equation}
There is only one primary field living on the $\mathbb{I}$-boundary,
the identity field $\mathbb{I}$ of weight $0$, while on the $\Phi$-boundary,
beside the identity field, there is an other primary, $\phi$, of
the weight $-\frac{1}{5}$. There are nontrivial boundary fields of
weight $-\frac{1}{5}$ interpolating the different boundary conditions,
denoted by $\psi$ and $\psi^{\dagger}$ and the Hilbert space of
the interpolating fields is $\mathcal{H}_{\psi}=\mathcal{H}_{\psi^{\dagger}}=V_{h}$.
\footnote{The field $\psi$ changes the boundary condition from $\phi$ to $\mathbb{I}$,
while $\psi^{\dagger}$ does the other way around.%
}

The boundary scaling Lee-Yang model is an integrable massive perturbation
of the conformal boundary Lee-Yang model. It allows a boundary parameter
\cite{Dorey:1997yg} 
\begin{equation}
S_{\Phi}(\lambda,\lambda_{b})=S_{\Phi}+\lambda\int\limits _{-\infty}^{\infty}dy\int\limits _{-\infty}^{0}dx\,\phi(x,y)+\lambda_{b}\int\limits _{-\infty}^{\infty}dy\,\varphi(y),
\end{equation}
where $S_{\Phi}$ denotes the action for the Lee-Yang model with the
$\Phi$-boundary condition imposed at $x=0$, and $\lambda$, $\lambda_{b}$
denote the bulk and boundary couplings, respectively. The action $S_{\mathbb{I}}(\lambda)$
of the perturbed theory with the identity boundary is similar, except
the boundary perturbation is missing. 

For $\lambda>0$ the perturbed theory is a massive scattering theory
having only a single particle type of mass $m(\lambda)$ with the
$S$ matrix \cite{Cardy:1990pc}: 
\begin{equation}
S(\theta)=\frac{\sinh\theta+i\sin\frac{\pi}{3}}{\sinh\theta-i\sin\frac{\pi}{3}}=-\left(\frac{1}{3}\right)\left(\frac{2}{3}\right)
\end{equation}
where we used the block notation introduced in (\ref{eq:blockdef}).
The pole at $\theta=\frac{2\pi i}{3}$ indicates that the particle
appears as a bound state of itself and such that the 3-particle coupling
is $\Gamma=i\sqrt{2\sqrt{3}}$. The mass of the Lee-Yang particle
as function of the perturbation parameter is given as
\begin{equation}
m(\lambda)=\kappa\lambda^{5/12}\quad,\qquad\kappa=\frac{2^{\frac{19}{5}}\sqrt{\pi}}{5^{\frac{5}{16}}}\frac{\left(\Gamma\left(\frac{3}{5}\right)\Gamma\left(\frac{4}{5}\right)\right)^{\frac{5}{12}}}{\Gamma\left(\frac{2}{3}\right)\Gamma\left(\frac{5}{6}\right)}.
\end{equation}

In the case of the $\textrm{\ensuremath{\mathbb{I}}}$ boundary the
reflection amplitude is 
\begin{equation}
R^{\mathbb{I}}(\theta)=\left(\frac{1}{2}\right)\left(\frac{1}{6}\right)\left(-\frac{2}{3}\right)
\end{equation}
which exhibits a pole at $i\frac{\pi}{2}$ with residue $g_{\mathbb{I}}=-2i\sqrt{(2\sqrt{3}-3)}$.
This shows that the $\mathbb{I}$ boundary can emit a virtual particle
with zero energy but there are no bound-states on this boundary.

The reflection factor of the $\Phi$-boundary depends on the strength
of the boundary coupling constant $\lambda_{b}$ as \cite{Dorey:1997yg}
\begin{equation}
R^{\Phi}(\theta)=R^{\mathbb{I}}(\theta)R_{\phi}(\theta)\quad,\quad R_{\phi}(\theta)=S(\theta-\theta_{0})S(\theta+\theta_{0})\quad,\quad\theta_{0}=i\pi\frac{3-b}{6},
\end{equation}
where the dimensionless parameter $b$ is related to the dimensionful
$\lambda_{b}$ as 
\begin{equation}
\lambda_{b}(b)=\sin\left(\bigl(b+\frac{1}{2}\bigr)\frac{\pi}{5}\right)m(\lambda)^{6/5}\lambda_{crit}\ ,\quad\lambda_{crit}=-\pi^{\frac{3}{5}}2^{\frac{4}{5}}5^{\frac{1}{4}}\frac{\sin\frac{2\pi}{5}}{\sqrt{\Gamma(\frac{3}{5})\Gamma(\frac{4}{5})}}\left(\frac{\Gamma(\frac{2}{3})}{\Gamma(\frac{1}{6})}\right)^{\frac{6}{5}}.
\end{equation}
The fundamental range of the parameter $b$ is $[-3,2]$ and we have
no boundary bound-state only in the region $b\in[-3,-1]$. This boundary
reflection factor can also emit a virtual zero energy particle with
amplitude 
\begin{equation}
g_{\Phi}\left(b\right)=\frac{\cosh\theta_{0}+\sin\frac{\pi}{3}}{\cosh\theta_{0}-\sin\frac{\pi}{3}}g_{\mathbb{I}}
\end{equation}
Note that $R^{\mathbb{I}}(\theta)$ is identical to $R^{\Phi}(\theta)$
at $b=0$ and so both have a pole at $\theta=\frac{i\pi}{2}$ coming
from the $\left(\frac{1}{2}\right)$ block, but their $g$ factors
differ in a sign \cite{Dorey:1998kt}. We also note that the $\Phi$-boundary
can be obtained by placing an integrable defect with transmission
factor $T(\theta)$ in front of the identity boundary
\begin{equation}
R^{\Phi}(\theta)=T_{-}(\theta)R^{\mathbb{I}}(\theta)T_{+}(\theta)
\end{equation}
In particular, the transmission factor satisfies $T_{\mp}(\theta)=S(\theta\mp\theta_{0})$,
thus it can be interpreted as an imaginary momentum bound particle.
This, however, does not mean that the $\Phi$-boundary is a boundary
bound-state as the $\mathbb{I}$ boundary has no bound-states. 

In the following we consider the situation in which we have the analogue
of $S_{\Phi}(\lambda,\lambda_{b})$ for $t<0$ and $S_{\mathbb{I}}(\lambda)$
for $t>0$, (or the other way around). The change in the boundary
condition is implemented by inserting the off-critical versions of
$\psi$ or $\psi^{\dagger}$ or their descendants and we analyze the
form factors of these operators. We use the general parametrization
\begin{equation}
F_{n}^{\mathcal{O}_{\beta\alpha}}(\theta_{1},\theta_{2},\dots,\theta_{n})=\left\langle \mathcal{O}_{\beta\alpha}\right\rangle H_{n}^{\beta\alpha}\prod_{i=1}^{n}\frac{r^{\beta\alpha}(\theta_{i})}{y_{i}}\prod_{i<j}\frac{f(\theta_{i}-\theta_{j})f(\theta_{i}+\theta_{j})}{(y_{i}+y_{j})}Q_{n}^{\mathcal{O}_{\beta\alpha}}(y_{1},y_{2}\dots,y_{n}).
\end{equation}
where we explicitly spelled out which quantities depend only on the
various boundary conditions and which depend on the operator itself.
From now on we will omit the operator if it does not lead to any confusion.

The minimal bulk two particle form factor, which has only a single
zero at $\theta=0$ and a pole at $\theta=\frac{2\pi i}{3}$ in the
strip $0\leq\Im{\rm m}(\theta)<\pi$, has the form \cite{Zamolodchikov:1990bk}:
\begin{equation}
f(\theta)=\frac{y-2}{y+1}v(i\pi-\theta)v(-i\pi+\theta)\quad,\quad y=e^{\theta}+e^{-\theta}
\end{equation}
 where 
\begin{equation}
v(\theta)=\exp\left\{ 2\int_{0}^{\infty}\frac{dt}{t}e^{i\frac{\theta t}{\pi}}\frac{\sinh\frac{t}{2}\sinh\frac{t}{3}\sinh\frac{t}{6}}{\sinh^{2}t}\right\} .
\end{equation}
It satisfies the important identities
\begin{equation}
f\left(\theta\right)f\left(\theta+i\pi\right)=\frac{\sinh\theta}{\sinh\theta-i\sin\frac{\pi}{3}}\quad,\qquad\frac{f\left(\theta+\frac{i\pi}{3}\right)f\left(\theta-\frac{i\pi}{3}\right)}{f(\theta)}=\frac{\cosh\theta+1/2}{\cosh\theta+1}.
\end{equation}

The one-particle minimal boundary changing form factor is parametrized
as
\begin{equation}
r^{\mathbb{I}\Phi}(\theta)=h^{\Phi}(\theta)h^{\mathbb{I}}(i\pi-\theta)=r^{\mathbb{II}}(\theta)r_{\phi}(\theta)\label{eq:r_id_phi_param}
\end{equation}
where 
\begin{equation}
r^{\mathbb{II}}(\theta)=4i\sinh\theta\exp\left\{ \int_{0}^{\infty}\frac{dt}{t}\frac{\sinh(t)-\cosh\left(\frac{it}{2}-\frac{\theta t}{\pi}\right)\left(\sinh\frac{5t}{6}+\sinh\frac{t}{2}-\sinh\frac{t}{3}\right)}{\sinh\frac{t}{2}\,\sinh t}\right\} 
\end{equation}
is the minimal form factor for the identity boundary condition. This
representation is valid on the strip $0\leq\Im{\rm m}(\theta)\leq\pi$
and can be extended by analytic continuation outside this region.
The identity boundary reflection factor satisfies

\begin{eqnarray}
r^{\mathbb{II}}(i\pi+\theta)r^{\mathbb{II}}(\theta)f(i\pi-2\theta) & = & y^{2}(y^{2}-4)\nonumber \\
\frac{r^{\mathbb{II}}(\theta+\frac{i\pi}{3})r^{\mathbb{II}}(\theta-\frac{i\pi}{3})}{r^{\mathbb{II}}(\theta)}f(2\theta) & = & y^{2}-3\nonumber \\
\frac{r^{\mathbb{II}}(\frac{i\pi}{2})}{v(0)} & = & 4\left(\sqrt{3}-3\right)
\end{eqnarray}
From the parametrization (\ref{eq:r_id_phi_param}) follows that $r_{\phi}$
satisfies
\begin{equation}
r_{\phi}(\theta)=R_{\phi}(\theta)r_{\phi}(-\theta)\quad,\qquad r_{\phi}(i\pi-\theta)=r_{\phi}(i\pi+\theta)
\end{equation}
and has no zeros or poles in the physical strip, thus the Karowski-Weisz
theorem implies 
\begin{equation}
\hspace{-0.24cm}r_{\phi}(\theta)=\mathcal{N}\exp\Big\{2\int_{0}^{\infty}\frac{dt}{t}\frac{\sinh\frac{b+1}{6}t+\sinh\frac{b-1}{6}t-\sinh\frac{b+7}{6}t-\sinh\frac{b+5}{6}t}{\sinh^{2}t}\sin^{2}\Big(\frac{i\pi-\theta}{2\pi}t\Big)\Big\}.
\end{equation}
This representation of $r_{\phi}$ is valid only for $b\in[-3,-1]$
and $0\leq\Im{\rm m}(\theta)\leq2\pi$, and defined by analytic continuation
outside this domain. If chose the normalization 
\begin{equation}
\mathcal{N}=-\frac{1}{4}\exp\left\{ 2\int_{0}^{\infty}\frac{dt}{t}\frac{\cosh\left(\frac{b+3}{6}t\right)\left[\sinh\frac{t}{3}+\sinh\frac{2t}{3}\right]-\sinh(t)}{\sinh^{2}(t)}\right\} 
\end{equation}
then $r_{\phi}(\theta)$ satisfies
\begin{eqnarray}
r_{\phi}(\theta)r_{\phi}(\theta+i\pi) & = & \frac{1}{(y_{0}-y_{-})(y_{0}+y_{+})}\\
\frac{r_{\phi}(\theta+\frac{i\pi}{3})r_{\phi}(\theta-\frac{i\pi}{3})}{r_{\phi}(\theta)} & = & \frac{1}{y+y_{0}}\\
r_{\phi}\left(\frac{i\pi}{2}\right) & = & \frac{1}{y_{0}-\sqrt{3}}
\end{eqnarray}
with
\begin{equation}
y_{+}=\omega e^{\theta}+\omega^{-1}e^{-\theta}\quad,\quad y_{-}=\omega e^{-\theta}+\omega^{-1}e^{\theta}\quad,\quad y_{0}=2\cosh\theta_{0}\quad,\quad\omega=e^{i\frac{\pi}{3}}.
\end{equation}

The minimal one particle boundary changing form factor corresponding
to $\alpha=\mathbb{I}$ and $\beta=\Phi$ is given as
\begin{equation}
r^{\Phi\mathbb{I}}(\theta)=-h^{\mathbb{I}}(\theta)h^{\Phi}(i\pi-\theta)=-r^{\mathbb{II}}(\theta)r_{\phi}(i\pi-\theta)
\end{equation}
where we defined an extra sign into $r^{\Phi\mathbb{I}}$ for later
convenience. 

By choosing the normalization $H_{n}^{\mathbb{I}\Phi}=H_{n}^{\Phi\mathbb{I}}=\left(\frac{i\sqrt[4]{3}}{v\left(0\right)\sqrt{2}}\right)^{n}$
the recursion relations for the polynomials become
\begin{eqnarray}
Q_{n+2}^{\beta\alpha}\left(y_{+},y_{-},y_{1},\dots,y_{n}\right) & = & D_{n}^{\beta\alpha}\left(y\vert y_{1},\dots,y_{n}\right)Q_{n+1}^{\beta\alpha}\left(y,y_{1},\dots,y_{n}\right)\label{eq:DynRec}\\
Q_{n+2}^{\beta\alpha}\left(y,-y,y_{1},\dots,y_{n}\right) & = & P_{n}^{\beta\alpha}\left(y\vert y_{1},\dots,y_{n}\right)Q_{n}^{\beta\alpha}\left(y_{1},\dots,y_{n}\right)\label{eq:KinRec}\\
Q_{n+1}^{\beta\alpha}\left(0,y_{1},\dots,y_{n}\right) & = & B_{n}^{\beta\alpha}\left(y_{1},\dots,y_{n}\right)Q_{n}^{\beta\alpha}\left(y_{1},\dots,y_{n}\right)\label{eq:BdryRec}
\end{eqnarray}
with
\begin{align}
P_{n}^{\mathbb{I}\Phi}(y\vert y_{1},\dots,y_{n}) & =\hspace{-0.3cm} & P_{n+1}^{\mathbb{II}}(y\vert y_{0},y_{1},\dots,y_{n}) & ,\; & P_{n}^{\Phi\mathbb{I}}(y\vert y_{1},\dots,y_{n}) & =P_{n+1}^{\mathbb{II}}(y\vert-y_{0},y_{1},\dots,y_{n})\\
D_{n}^{\mathbb{I}\Phi}(y\vert y_{1},\dots,y_{n}) & =\hspace{-0.3cm} & D_{n+1}^{\mathbb{II}}(y\vert y_{0},y_{1},\dots,y_{n}) & ,\; & D_{n}^{\Phi\mathbb{I}}(y\vert y_{1},\dots,y_{n}) & =D_{n+1}^{\mathbb{II}}(y\vert-y_{0},y_{1},\dots,y_{n})\\
B_{n}^{\mathbb{I}\Phi}(y_{1},\dots,y_{n}) & =\hspace{-0.3cm} & B_{n+1}^{\mathbb{II}}(y_{0},y_{1},\dots,y_{n}) & ,\; & B_{n}^{\Phi\mathbb{I}}(y_{1},\dots,y_{n}) & =B_{n+1}^{\mathbb{II}}(-y_{0},y_{1},\dots,y_{n})
\end{align}
and

\begin{eqnarray}
D_{n}^{\mathbb{II}}\left(y\vert y_{1},\dots,y_{n}\right) & = & \prod_{i=1}^{n}\left(y+y_{i}\right)\label{eq:DynPoly}\\
P_{n}^{\mathbb{II}}\left(y\vert y_{1},\dots,y_{n}\right) & = & \frac{\prod_{i=1}^{n}\left(y_{i}-y_{-}\right)\left(y_{i}+y_{+}\right)-\prod_{i=1}^{n}\left(y_{i}+y_{-}\right)\left(y_{i}-y_{+}\right)}{2\left(y_{+}-y_{-}\right)}\label{eq:KinPoly}\\
B_{n}^{\mathbb{II}}\left(y_{1},\dots,y_{n}\right) & = & \frac{\prod_{i=1}^{n}\left(y_{i}+\sqrt{3}\right)-\prod_{i=1}^{n}\left(y_{i}-\sqrt{3}\right)}{2\sqrt{3}}\label{eq:BdryPoly}
\end{eqnarray}

\subsubsection{Form factors of the primary boundary changing fields}

The form factor recurrence relations (\ref{eq:DynRec},\ref{eq:KinRec},\ref{eq:BdryRec})
have many sets of solution. We expect that the ones with the mildest
ultraviolet behaviour correspond to the off-critical versions of the
primary boundary changing fields, $\psi$ and $\psi^{\dagger}$, with
the appropriate boundaries. Observe that the recursion relations are
exactly the same that we would get for $Q_{n+1}^{\mathbb{II}}(\pm y_{0},y_{1},\dots,y_{n})$
for the boundary form factors on the $\mathbb{I}$ boundary condition
\cite{Bajnok:2006ze}. So that one may expect to get the $Q$-polynomials
of the fields $\psi$ and $\psi^{\dagger}$ from the polynomials corresponding
to the off-critical version of the energy-momentum tensor with identity
boundary, which is the lowest-lying solution in the that case, by
setting $y_{n+1}\rightarrow\pm y_{0}$. However, there is an essential
difference between the case of boundary changing operators and the
identity boundary case, namely in the latter case the one-particle
form factor does not have the boundary kinematical pole while in the
former case it does. The vanishing of the residue of the boundary
kinematical pole requires $Q_{1}^{\mathbb{II}}(0)=0$ thus $Q_{0}^{\mathbb{II}}=0$
for all operators living on the identity boundary, but we expect $Q_{0}^{\psi}=Q_{0}^{\psi^{\dagger}}=1$. 

The solution for the off-critical energy-momentum tensor with identity
boundary condition was determined in \cite{Hollo:2014vpa} and reads
as
\begin{equation}
Q_{1}^{T}=\sigma_{1}^{(1)}\quad;\qquad Q_{2}^{T}=\sigma_{1}^{(2)}\quad;\qquad Q_{3}^{T}=\left(\sigma_{1}^{(3)}\right)^{2}\quad;\qquad Q_{n}^{T}=\left(\sigma_{1}^{(n)}\right)^{2}\det\Xi^{(n)}
\end{equation}
for $n\geq4$ where the $(n-3)\times(n-3)$ matrix function is defined
as
\begin{equation}
\Xi_{ij}^{(n)}=\sum_{k\in\mathbb{Z}}3^{k}\binom{i-j+k}{k}\sigma_{3j-2i+1-2k}^{(n)}\qquad,\qquad1\leq i,j\leq n-3
\end{equation}

However it is still possible the generate the form factor solutions
for the boundary changing primaries from the solution for the energy-momentum
tensor. Let us observe that the $\sigma_{1}^{(n)}$ symmetric polynomial,
introduced in (\ref{eq:symm_polys}), is a zero mode of the recurrence
equations (\ref{eq:DynRec},\ref{eq:KinRec},\ref{eq:BdryRec}), i.e.
\[
\sigma_{1}(y_{+},y_{-},y_{1}\dots y_{n})=\sigma_{1}(y,y_{1},\dots,y_{n})\quad,\qquad\sigma_{1}(-y,y,y_{1},\dots,y_{n})=\sigma_{1}(y_{1},\dots,y_{n})
\]
\begin{equation}
\sigma_{1}(0,y_{1},\dots,y_{n})=\sigma_{1}(y_{1},\dots,y_{n})
\end{equation}
thus every solution can be multiplied or, if divisible, divided by
$\sigma_{1}$! Dividing the $Q_{n+1}^{T}$ polynomial, corresponding
to the energy-momentum tensor in the identity boundary case, by $\sigma_{1}^{(n+1)}$
and evaluating it at $y_{n+1}=\pm y_{0}$ will generate the solution
for $\psi$ and $\psi^{\dagger}$ with the appropriate initial conditions,
$Q_{0}^{\psi}=Q_{0}^{\psi^{\dagger}}=1$,
\begin{equation}
Q_{n}^{\psi}(y_{1},\dots,y_{n})=\left.\frac{Q_{n+1}^{T}}{\sigma_{1}^{(n+1)}}\right|_{(y_{0},y_{1},\dots,y_{n})}\qquad,\qquad Q_{n}^{\psi^{\dagger}}(y_{1},\dots,y_{n})=\left.\frac{Q_{n+1}^{T}}{\sigma_{n+1}^{(1)}}\right|_{(-y_{0},y_{1},\dots,y_{n})}
\end{equation}

\subsubsection{Two point functions of boundary operators and their UV limits }

Let us consider the off-critical two-point functions of the Euclidean
version of the model 
\begin{equation}
\left\langle \varphi_{1}(r)\varphi_{2}(0)\right\rangle 
\end{equation}
where $\varphi_{i}$ ($i=1,2$) is one of the off-critical version
of the boundary fields $\phi$, $\psi$ and $\psi^{\dagger}$ compatible
with the corresponding boundary conditions. The two point function
can be computed via its spectral representation
\begin{equation}
\langle0\vert\varphi_{1}(r)\varphi_{2}(0)\vert0\rangle=\sum_{n=0}^{\infty}\int_{\theta_{1}>\dots>\theta_{n}>0}\frac{d\theta_{1}}{2\pi}\dots\frac{d\theta_{n}}{2\pi}\,\, e^{-r\Delta E_{{\rm bdry}}^{\varphi_{1}}-mr\sum_{i}\cosh\theta_{i}}F_{n}^{\varphi_{1}}F_{n}^{\varphi_{2}+}\label{eq:Spectr_repr}
\end{equation}
where
\begin{eqnarray}
F_{n}^{\varphi_{1}} & = & \langle0\vert\varphi_{1}(0)\vert\theta_{1},\dots,\theta_{n}\rangle_{in}=F_{n}^{\varphi_{1}}(\theta_{1},\dots,\theta_{n})\nonumber \\
F_{n}^{\varphi_{2}+} & = & \phantom{i}_{in}\langle\theta_{1},\dots,\theta_{n}\vert\varphi_{2}(0)\vert0\rangle=F_{n}^{\varphi_{2}}(i\pi+\theta_{n},\dots,i\pi+\theta_{1})
\end{eqnarray}
and $\Delta E_{{\rm bdry}}^{\varphi_{1}}$ is the difference of the
boundary energies of the boundary conditions in between $\varphi_{1}$
interpolates,
\begin{equation}
\Delta E_{{\rm bdry}}^{\phi}=0\quad,\qquad\Delta E_{{\rm bdry}}^{\psi^{\dagger}}=-\Delta E_{{\rm bdry}}^{\psi}=\frac{y_{0}}{2}
\end{equation}
Truncation of the series (\ref{eq:Spectr_repr}) up to two particle
term gives a good approximation even for small separation which can
be compared to the CFT prediction. Assuming that there is a one-to-one
correspondence between the field content of the perturbed theory and
the CFT (apart form some additive renormalization constant \cite{Zamolodchikov:1990bk})
we can use the operator product expansion of the CFT
\begin{equation}
\varphi_{1}(r)\varphi_{2}(0)\sim\sum_{j}\frac{C_{12}^{j}\varphi_{j}}{\left|r\right|^{h_{1}+h_{2}-h_{j}}}\label{eq:OPE_series}
\end{equation}
where the sum runs over all the boundary fields, and $h_{j}$ denotes
the weights of the fields. Choosing $\varphi_{1}$ and $\varphi_{2}$
to be primaries and keeping the leading contributions in (\ref{eq:OPE_series})
with the lowest weights, i.e. the primaries appearing in the OPE of
$\varphi_{1}$ and $\varphi_{2}$, we get a good approximation of
the short distance behavior of the two point functions. The OPEs of
interest are
\begin{equation}
\phi(z)\psi^{\dagger}(w)=C_{\phi\psi^{\dagger}}^{\psi^{\dagger}}\left|z-w\right|^{1/5}\psi^{\dagger}(w)+\dots\quad;\quad\psi(z)\phi(w)=C_{\psi\phi}^{\psi}\left|z-w\right|^{1/5}\psi(w)+\dots
\end{equation}
with the structure constants
\begin{equation}
C_{\phi\psi^{\dagger}}^{\psi^{\dagger}}=C_{\psi\phi}^{\psi}=-\sqrt{\frac{2}{1+\sqrt{5}}}\sqrt{\frac{\Gamma\left(\frac{1}{5}\right)\Gamma\left(\frac{6}{5}\right)}{\Gamma\left(\frac{3}{5}\right)\Gamma\left(\frac{4}{5}\right)}}
\end{equation}
As the exact vacuum expectation values of the boundary (changing)
fields are only known for $\phi$ \cite{Dorey:2000eh}
\begin{equation}
\left\langle \phi\right\rangle =-\frac{5}{6\left|\lambda_{crit}\right|}\frac{\cos\left(b\pi/6\right)}{\cos\left(\pi(b+1/2)/5\right)}m^{-\frac{1}{5}}
\end{equation}
we will consider the normalized two point functions%
\footnote{Here the ground state expectation value is meant as the matrix element
between the lowest energy states corresponding to the various boundary
condition. It can either be the the highest weight state of the $V_{0}$
module, $\vert0\rangle$, in the identity boundary case or the highest
weight state of the $V_{h}$ module, $\vert\phi\rangle$, in the $\Phi$-boundary
case.%
}
\begin{equation}
\frac{\left\langle \psi(r)\phi(0)\right\rangle }{\left\langle \psi\right\rangle \left\langle \phi\right\rangle }=\frac{C_{\psi\phi}^{\psi}}{\left\langle \phi\right\rangle }(mr)^{1/5}+\dots\quad;\qquad\frac{\left\langle \phi(r)\psi^{\dagger}(0)\right\rangle }{\left\langle \phi\right\rangle \left\langle \psi^{\dagger}\right\rangle }=\frac{C_{\phi\psi^{\dagger}}^{\psi^{\dagger}}}{\left\langle \phi\right\rangle }(mr)^{1/5}+\dots
\end{equation}
As the form factors are also proportional to the vacuum expectation
values of the fields, it drops out in the normalized version. 

For the numerical implementation of the truncated form factor series
we need the form factors of the boundary field $\phi$. They are parametrized
as
\begin{equation}
F_{n}^{\phi}(\theta_{1},\dots,\theta_{n})=\left\langle \phi\right\rangle H_{n}^{\Phi\Phi}\prod_{i=1}^{n}\frac{r^{\Phi\Phi}(\theta_{i})}{y_{i}}\prod_{i<j}\frac{f(\theta_{i}-\theta_{j})f(\theta_{i}+\theta_{j})}{y_{i}+y_{j}}Q_{n}^{\phi}(y_{1},\dots,y_{n})
\end{equation}
with
\begin{equation}
r^{\Phi\Phi}(\theta)=\frac{1}{4\left(\sinh\theta-i\sin\pi\frac{b-1}{6}\right)\left(\sinh\theta-i\sin\pi\frac{b+1}{6}\right)}r^{\mathbb{II}}(\theta)
\end{equation}
and $H_{n}^{\Phi\Phi}=\left(\frac{i\sqrt[4]{3}}{v\left(0\right)\sqrt{2}}\right)^{n}$.
The polynomials $Q_{n}^{\phi}$ are calculated explicitly in \cite{Hollo:2014vpa},
we only need the first few of them, which are
\begin{equation}
Q_{1}^{\phi}=\sigma_{1}^{(1)}\quad,\qquad Q_{2}^{\phi}=\sigma_{1}^{(2)}\left(\sigma_{2}^{(2)}+3-y_{0}^{2}\right).
\end{equation}

\begin{figure}
\begin{centering}
\includegraphics[width=14cm]{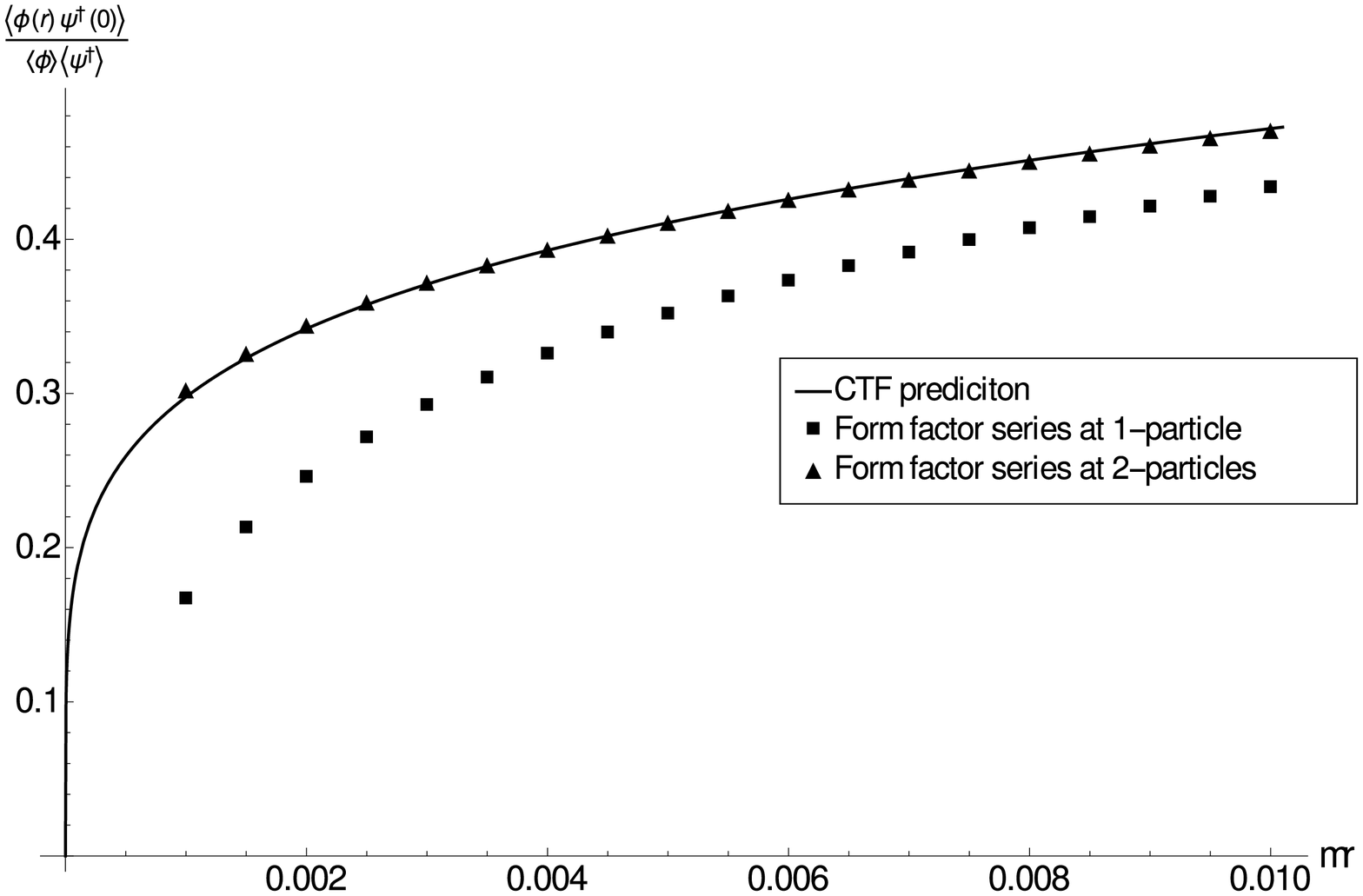}
\par\end{centering}

\protect\caption{\label{fig:2pt}The normalized $\langle\phi(r)\psi^{\dagger}(0)\rangle$
two point function at $b=-2$. }
\end{figure}

We numerically calculated the one- and two-particle contributions
to the normalized two point functions and plotted against the CFT
prediction, shown in Figure \ref{fig:2pt}, which shows a good agreement.
This is a solid confirmation of our solutions for the form factors
of $\psi$ and $\psi^{\dagger}$.
\begin{figure}
\begin{centering}
\includegraphics[width=14cm]{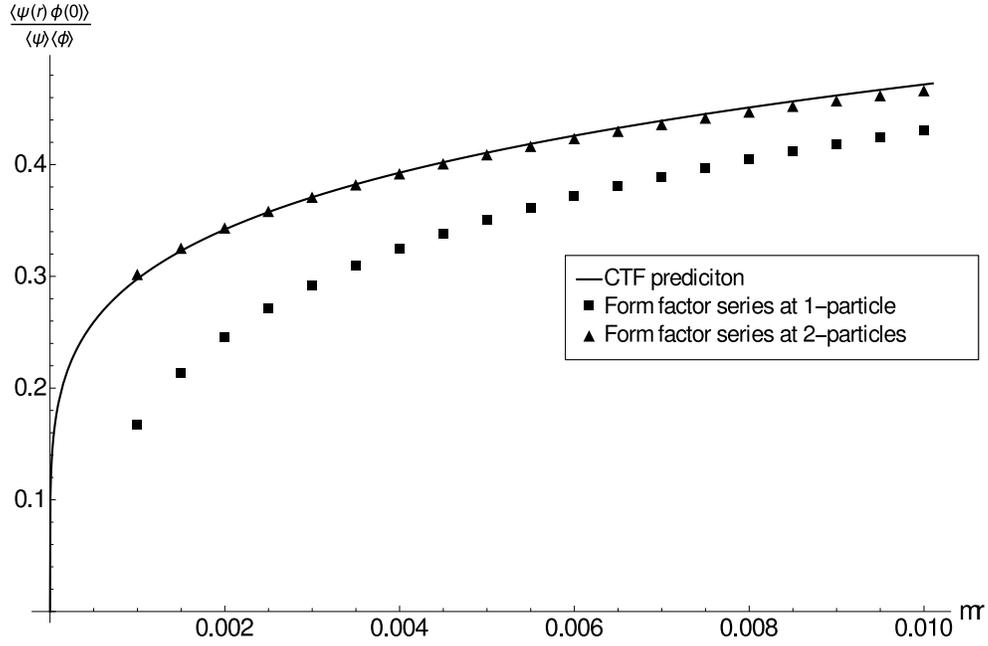}
\par\end{centering}

\protect\caption{\label{fig:2pt-1}The normalized $\langle\psi(r)\phi(0)\rangle$ two
point function at $b=-2$. }
\end{figure}

\subsubsection{Classification of the form factor solutions}

In this subsection we classify the polynomial solutions of the recursion
relations following \cite{Koubek:1993ke,Koubek:1994gk,Koubek:1994zp,Szots:2007jq}.
The asymptotic degree of a form factor solution is defined as 
\begin{equation}
\lim_{\Lambda\to\infty}F_{n}^{\beta\alpha}(\theta_{1}+\Lambda,\dots,\theta_{n}+\Lambda)=e^{x_{n}\Lambda}+\dots
\end{equation}
and is in one-to-one correspondence with the UV scaling dimension
of the operator. Using the parametrization of the form factors together
with their asymptotic behaviour their degree turns out to be
\begin{equation}
x_{n}=\deg Q_{n}-\frac{n(n-1)}{2}
\end{equation}
The form factor of each boundary changing operator starts at a given
particle number and all form factors with more particles are uniquely
determined from this first. Such family of solutions, which defines
the operator, is called the form factor tower and the scaling dimension
can be read off from the degree of the top of the tower. So far we
considered only the solution which starts at the first level and has
the mildest asymptotic growth, but there are also other solutions.
They correspond to the so-called kernel solutions and can start at
any higher level. An $n^{\mathrm{th}}$ level kernel solution is defined
as a polynomial of $n$ variables\emph{ }whose value is zero at the
positions of all the singularity axioms.\emph{ }In the case of the
boundary Lee-Yang model they are given as 
\begin{equation}
Q_{n}=\sigma_{k_{1}}^{\left(n\right)}\dots\sigma_{k_{l}}^{\left(n\right)}K_{n}\quad;\quad K_{n}=\prod_{1\leq i<j\leq n}\left(y_{i}+y_{j}\right)\prod_{1\leq i<j\leq n}\left(y_{i}^{2}+y_{i}y_{j}+y_{j}^{2}-3\right)\prod_{i=1}^{n}y_{i}
\end{equation}
where $0<k_{1}\leq k_{2}\leq\dots\leq k_{l}\leq n$. The corresponding
form factor has degree
\begin{equation}
x_{n}=k_{1}+\dots+k_{l}+n^{2}
\end{equation}
Formally we can consider the fundamental solution corresponding to
$K_{1}=1$. Its descendant $\sigma_{1}^{n}K_{1}$ is nothing but its
$n^{th}$ derivative. The generating function of all the solutions
is 
\begin{equation}
1+\sum_{n=1}^{\infty}\sum_{lm=0}^{\infty}P(m\vert n)q^{l+n^{2}}
\end{equation}
where $P(m\vert n)$ denotes the number of partitions of the number
$m$ such that none of summands is greater than $n$, and the extra
$1$ corresponds to $K_{1}$. Using 
\begin{equation}
\sum_{m=0}^{\infty}P(m\vert n)q^{m}=\prod_{i=1}^{n}(1-q^{i})^{-1}
\end{equation}
and the Rogers-Ramanujan identity we can write
\begin{equation}
1+\sum_{n=1}^{\infty}\sum_{l=0}^{\infty}P(l\vert n)q^{l+n^{2}}=1+\sum_{n=1}^{\infty}\frac{q^{n^{2}}}{\prod_{l=1}^{n}(1-q^{l})}=\prod_{n=0}^{\infty}\frac{1}{(1-q^{5n+1})(1-q^{5n+4})}=\tilde{\chi}_{-\frac{1}{5}}
\end{equation}
which is the truncated character of the $h=-\frac{1}{5}$ representation.
Thus at each level we found exactly the same number of form factor
solutions as many state exist at that conformal level. As there is
an isomorphism between states and local boundary changing operators
in a CFT we can see that there is a one-to-one correspondence between
the form factor solutions and local boundary changing operators.

\section{Conclusion\label{sec:Conclusion}}

In this paper we established the form factor bootstrap program for
boundary condition changing operators in integrable models. Our proposal
fills some gap as, although the complete set of form factor axioms
were known for a long time for bulk \cite{Smirnov:1992vz,Babujian:1998uw},
boundary \cite{Bajnok:2006ze} and defect \cite{Bajnok:2009hp} models,
and also for some non-local operator insertions \cite{Bajnok:2015hla},
the complete set of axioms for the form factors of local boundary
changing operators were missing. We have tested the consistency of
the form factor axioms and presented the general procedure to determine
their solutions. 

The first step of the method is the calculation of the one-particle
minimal form factor. Whenever the reflection factors of the two boundaries
can be written as a product of blocks (\ref{eq:blockdef}), the ingredients
of the minimal solution are granted by the theorem of Karowski and
Weisz \cite{Karowski:1978vz}. Then, a general multiparticle form
factor can be parametrized in terms of the minimal boundary form factor
and the bulk two-particle minimal form factor, which automatically
satisfies some of the axioms. This parametrization includes a polynomial
factor, and the rest of the form factor axioms give restrictive recursive
relations connecting these polynomials. There is a one-to-one correspondence
between the families of solutions of the recurrence relations and
the operator content of the model \cite{Koubek:1993ke,Koubek:1994gk,Koubek:1994zp,Szots:2007jq}.

In the pioneering paper \cite{Lesage:1998hh} the authors analyzed
in detail the free massive fermion and the sinh-Gordon model. Here,
we analyzed two other models in detail. First, in the boundary condition
changing free boson theory, we solved the form factor bootstrap axioms.
If, at a moment, the boundary condition is changed, the vacuum of
the pre-quench system becomes an excited state of the post-quench
one. We presented the explicit relation of the two vacua involving
two kernel functions satisfying specific integral equations. We gave
the relation of these kernel functions to the one- and two-particle
form factors. When the boundary condition is changed from Neumann
to Dirichlet, we showed that the form factor bootstrap solutions indeed
satisfy these integral equations. It would be interesting to prove
that it also holds for the generic case. 

A finite volume analysis was presented in the case when the boundary
condition is switched from Neumann to Dirichlet, by introducing a
second boundary at $x=-L$ with Neumann boundary condition. In fact,
the boundary condition of the new boundary is not relevant as we take
the $L\rightarrow\infty$ limit at the end. The before and after quench
boson creation and annihilation operators, as in the infinite volume
case, are related to each other by a Bogoliubov-type transformation.
By hermitian conjugation we can flip back the outgoing Dirichlet states
and the new incoming states are now tensor products of two free boson
states. The vertex state is defined such that the overlap of an incoming
and an outgoing state before the flipping, i.e. the form factor of
the quench operator, is equal to the overlap of the flipped incoming
state and the vertex state. We parametrized the vertex state in terms
of the so-called Neumann coefficients, and the relations connecting
the creation and annihilation operators result restrictive equations
for the Neumann coefficients. A similar problem had been analyzed
in the context of the open-closed string vertex \cite{Lucietti:2003ki,Lucietti:2004wy}.
If we consider Dirichlet boundary condition on the open string than
the resulting equations for the string vertex can be mapped to our
equations for the Neumann coefficients, thus we could simply read
of the solutions. By definition, the vertex state contain all the
information of the form factors, thus by taking the $L\rightarrow\infty$
limit of the Neumann coefficients we could determine directly the
infinite volume form factors of the boundary changing operator. The
resulting functions coincide with the bootstrap prediction which confirms
the validity of our axioms.

We also considered the scaling Lee-Yang model. There are only two
integrable boundary condition, the identity boundary and the so-called
$\Phi$-boundary. We studied both the case when we switch from the
identity to the $\Phi$-boundary and the other way around. First,
we calculated the minimal boundary-changing one-particle form factors
and then we derived the recursive relations for the polynomials appearing
in the parametrization of the multiparticle form factors. These recurrence
equations turned out to be very similar to the ones for the (unquenched)
identity boundary \cite{Bajnok:2006ze}, whose solutions are known
\cite{Hollo:2014vpa}. We gave the explicit solutions for the form
factors corresponding to the boundary changing operators with the
mildest ultraviolet behaviour, i.e. the off-critical versions of the
conformal boundary changing primary fields. By analyzing the structure
of the recurrence relations, we found their common kernels. By counting
the kernel solutions we showed that there is a one-to-one correspondence
between the operator content of the theory and the towers of solutions
of the form factor axioms. Finally, we studied the two-point correlation
functions of a boundary and a boundary changing operator. Their spectral
series, truncated at two-particle level, give a good approximation
of the two-point function even in relatively small volume. We compared
this against the conformal field theory prediction, and we found a
good agreement. This supports the validity of our form factor solutions.

In the future it would be interesting to generalize the truncated
conformal space approach to describe boundary changing operators in
order to test our results, similarly how this check was carried out
for boundary form factors in \cite{Lencses:2011ab} and for defect
form factors in \cite{Bajnok:2013eaa}. 

Our framework is very general and can be directly used to calculate
the form factors of the boundary changing operators in other diagonal
models. The generalization of the program for non-diagonal theories
is also very interesting. 

From the quench problem point of view our result provides the exact
overlap of the pre-quench vacuum with all the post-quench states.
This result could be used to calculate interesting physical quantities
like correlation functions which can shed light on thermalization
or can characterize steady states.

\subsection*{Acknowledgments}

We thank Z. Laczkó for his collaboration at an early stage of this
research and J. Konczer for various discussions. We are grateful to
G. Takács and M. Kormos for their comments on the manuscript. We are
grateful to the Yukawa Institute for Theoretical Physics for their
hospitality where some part of the work was carried out. ZB and LH
were supported by a Lendület grant. LH has received funding from the
European Research Council (Programme Ideas ERC-2012-AdG 320769 AdS-CFT-solvable)
and from the Emberi Er\H{o}források Támogatáskezel\H{o} (NTP-EFÖ-P-15-0088).

\appendix

\section{Formal derivation of the axioms from the ZF algebra\label{sec:ZF_algebra}}

Here we present a formal derivation of our axioms from the Zamolodchikov-Faddeev
algebra%
\footnote{Similar consideration had been presented in \cite{Lesage:1998hh}.%
}. This algebra contain the exact operators $Z^{+}(\theta)$ and $Z(\theta)$
which create and annihilate particles. Formally they can be continued
for complex rapidities and the crossing transformation relates them
as 
\begin{equation}
Z(\theta)=Z^{+}(\theta+i\pi)
\end{equation}
These operators satisfy an exchange axiom including the exact scattering
matrix
\begin{equation}
Z^{+}(\theta_{1})Z^{+}(\theta_{2})=S(\theta_{1}-\theta_{2})Z^{+}(\theta_{2})Z^{+}(\theta_{1})+2\pi\delta(\theta_{1}-\theta_{2}-i\pi)\label{eq:bulk_ZF}
\end{equation}
such that the exchange of the creation and annihilation operators
contain the $\delta$ function, too. 

In the presence of the boundary we introduce the boundary operators:
\begin{equation}
\vert0\rangle^{\alpha}=B_{\alpha}^{+}\vert0\rangle\quad,\qquad\,^{\beta}\langle0\vert=\langle0\vert B_{\beta}
\end{equation}
such that 
\begin{equation}
Z^{+}(\theta)B_{\alpha}^{+}=R_{\alpha}(\theta)Z^{+}(-\theta)B_{\alpha}^{+}+2\pi\delta(\theta-\frac{i\pi}{2})\frac{g_{\alpha}}{2}B_{\alpha}^{+}
\end{equation}
and 
\begin{equation}
B_{\beta}Z(\theta)=B_{\beta}Z(-\theta)R_{\beta}(-\theta)+2\pi\delta(\theta+\frac{i\pi}{2})\frac{g_{\beta}}{2}B_{\beta}
\end{equation}

The form factor axioms can be derived from the representation 
\begin{equation}
F_{n}^{\mathcal{O}_{\beta\alpha}}(\theta_{1},\dots,\theta_{n})=\,\langle0\vert B_{\beta}\,\mathcal{O}_{\beta\alpha}(0)\, Z^{+}(\theta_{1})\dots Z^{+}(\theta_{n})B_{\alpha}^{+}\vert0\rangle
\end{equation}
by assuming 
\begin{equation}
[\mathcal{O}_{\beta\alpha}(0),Z^{+}(\theta)]=0
\end{equation}

\section{Changing the boundary condition from Neumann to Dirichlet\label{sec:Neumann_to_Dirichlet}}

In this Appendix we analyze a simplified situation in which the Neumann
boundary condition is changed to Dirichlet in the free boson theory.
As a start we recall the bootstrap solution of the problem and show
how it solves explicitly the constraints coming from the direct quantization.
In the direct quantization the creation and annihilation operators
of the two boundary conditions are related to each other by an infinite
dimensional linear transformation. As a consequence, the vacuum state
of the Neumann boundary condition is a complicated coherent state
for the Dirichlet boundary (\ref{eq:Vac_rel}), and the appearing
kernels, the solutions of (\ref{eq:K1_condition},\ref{eq:K2_condition}),
can be found by inverting an infinite dimensional matrix, $A_{kk'}$.
Although we cannot invert this matrix, we can show that the bootstrap
solution provides a solution for the kernels.

In order to find the solution directly we put the system into a finite
volume by introducing Neumann condition at the other end. We can map
this finite volume problem to the open closed string vertex problem
\cite{Lucietti:2003ki,Lucietti:2004wy} and the adopted solution in
the infinite volume limit indeed reproduces the bootstrap result.

\subsection{Bootstrap solution}

Let us specify the bootstrap solution of Section \ref{sub:Free_boson}
for the case when the Neumann boundary condition, labeled by $\alpha=+$
with reflection factor $R^{\alpha}(\theta)\equiv1$, is changed to
the Dirichlet boundary, labeled by $\beta=-$ with reflection factor
$R^{\beta}(\theta)\equiv-1$. This limiting case can be obtained from
the general considerations as the $\lambda^{\alpha}\to0$ and $\lambda^{\beta}\to\infty$
limits. First, we need to calculate the one particle minimal form
factor
\begin{equation}
r^{-+}(\theta)=h^{+}(\theta)h^{-}(i\pi-\theta)
\end{equation}
which turns out to be
\begin{equation}
h^{+}(\theta)=1\quad,\qquad h^{-}(\theta)=2\sinh\frac{\theta}{2}\quad,\qquad r^{-+}(\theta)=2\sinh\left(\frac{i\pi-\theta}{2}\right).
\end{equation}
We choose the normalization such that 
\begin{equation}
r^{-+}(\theta)r^{-+}(i\pi+\theta)=-2i\sinh\theta.
\end{equation}
The general $n$-particle form factor is parametrized as
\begin{equation}
F_{n}^{-+}(\theta_{1},\theta_{2},\dots,\theta_{n})=\mathcal{N}H_{n}G_{n}(y_{1},\dots,y_{n})\prod_{i=1}^{n}r^{-+}(\theta_{i})\quad;\qquad y_{i}=e^{\theta_{i}}+e^{-\theta_{i}}
\end{equation}
where $\mathcal{N}=\phantom{I}^{-}\langle0\vert0\rangle^{+}$ play
the role of the vacuum expectation value. The kinematical residue
equation 
\begin{equation}
-i\mathop{\textrm{Res}}_{\theta=\theta'}F_{n+2}^{-+}(\theta+i\pi,\theta',\theta_{1},\dots,\theta_{n})=-2F_{n}^{-+}(\theta_{1},\dots,\theta_{n})
\end{equation}
connects either the even or the odd particle form factors to each
other. The solution, starting with $G_{0}=1$ and $G_{1}(y)\equiv1$
is given by 
\begin{equation}
G_{n}=\frac{1}{y_{nn-1}}G_{n-2}+\mbox{perm}=\sum_{\mathrm{all\, pairings}}\frac{1}{\prod_{\mathrm{all\, pairs}(i,j)}y_{ij}}
\end{equation}
where $H_{2n}=(-2)^{n}$ and $y_{ij}=y_{i}+y_{j}$. Here we chose
a slightly different normalization for both $r^{-+}$ and $H_{2n}$
form the ones in Section \ref{sub:Free_boson}, but the form factors
are the same.

\subsection{Direct infinite volume calculation}

The expansion of the free boson field with the Neumann or Dirichlet
boundary conditions are

\begin{equation}
\Phi(x,t)=\begin{cases}
\int_{0}^{\infty}\tilde{dk}\left\{ a_{+}(k)e^{-i\omega(k)t}+a_{+}^{+}(k)e^{i\omega(k)t}\right\} \phi_{k}^{+}(x) & t<0\\
\int_{0}^{\infty}\tilde{dk}\left\{ a_{-}(k)e^{-i\omega(k)t}-a_{-}^{+}(k)e^{i\omega(k)t}\right\} \phi_{k}^{-}(x) & t>0
\end{cases}\ ,\ \phi_{k}^{\pm}(x)=e^{ikx}\pm e^{-ikx}\label{eq:Field}
\end{equation}
where the creation/annihilation operators are normalized as 
\begin{equation}
\left[a_{\pm}(k),a_{\pm}^{+}(k')\right]=4\pi\omega(k)\delta(k-k')\label{eq:N_D_mode_algebra}
\end{equation}
The modes are orthogonal with a given boundary condition (\ref{eq:Orthogonality})
and they form a complete system (\ref{eq:Completeness}), so each
basis can be expressed in terms of the other
\begin{equation}
\int_{-\infty}^{0}\phi_{k}^{\pm*}(x)\phi_{k'}^{\mp}(x)dx=2i\frac{(k+k')\mp(k-k')}{k^{2}-k'^{2}}=2i\frac{(k+k')\mp(k-k')}{\omega^{2}(k)-\omega^{2}(k')}\equiv A_{kk'}^{\pm\mp}
\end{equation}
As the quantum field, $\Phi$, and its conjugate momentum, $\partial_{t}\Phi=\Pi$,
is continuous in the bulk, we can relate the creation and annihilation
operators of different boundary conditions to each other. Projecting
$\Phi(x,t=0)$ and $\Pi(x,t=0)$ onto the modes and combining them
results 
\begin{eqnarray}
a_{+}(k) & = & \int_{0}^{\infty}\frac{i}{\pi}\frac{k'\, dk'}{\omega(k')}\left\{ \frac{a_{-}(k')}{\omega(k)-\omega(k')}-\frac{a_{-}^{+}(k')}{\omega(k)+\omega(k')}\right\} \nonumber \\
a_{-}(k) & = & k\int_{0}^{\infty}\frac{i}{\pi}\frac{dk'}{\omega(k')}\left\{ \frac{a_{+}(k')}{\omega(k)-\omega(k')}+\frac{a_{+}^{+}(k')}{\omega(k)+\omega(k')}\right\} 
\end{eqnarray}
These are nothing but infinite dimensional Bogliubov transformations.
The vacuum state of the Neumann boundary condition is a complicated
coherent state for the Dirichlet boundary condition (\ref{eq:Vac_rel}),
and we parametrize it as
\begin{equation}
\vert0\rangle^{+}=\mathcal{N}\left(1+\int_{0}^{\infty}\tilde{dk_{0}}K_{1}^{+-}(k_{0})a_{-}^{+}(k_{0})\right)\exp\left\{ \frac{1}{2}\iint_{0}^{\infty}\tilde{dk_{1}}\tilde{dk_{2}}K_{2}^{+-}(k_{1},k_{2})a_{-}^{+}(k_{1})a_{-}^{+}(k_{2})\right\} \vert0\rangle^{-}
\end{equation}
where $K_{2}^{+-}$ is symmetric in its arguments, and $\mathcal{N}=\phantom{I}^{-}\langle0\vert0\rangle^{+}$.
Now demanding $a_{+}\vert0\rangle^{+}=0$ constrains the form of the
$K_{1}^{+-}$ and $K_{2}^{+-}$ kernels, which are the solutions of
\begin{equation}
0=\int_{0}^{\infty}k'\tilde{dk'}\frac{1}{\omega(k)-\omega(k')}K_{1}^{+-}(k')\label{eq:K1+-}
\end{equation}
and
\begin{equation}
-\frac{k'}{\omega(k)+\omega(k')}-\int_{0}^{\infty}\tilde{dk_{1}}\frac{k_{1}}{\omega(k_{1})-\omega(k)}K_{2}^{+-}(k_{1},k')=0\label{eq:K2+-}
\end{equation}

Or, the other way around, we can express the Dirichlet vacuum with
Neumann
\begin{equation}
\vert0\rangle^{-}=\mathcal{N}^{*}\left(1+\int_{0}^{\infty}\tilde{dk_{0}}K_{1}^{-+}(k_{0})a_{+}^{+}(k_{0})\right)\exp\left\{ \frac{1}{2}\iint_{0}^{\infty}\tilde{dk_{1}}\tilde{dk_{2}}K_{2}^{-+}(k_{1},k_{2})a_{+}^{+}(k_{1})a_{+}^{+}(k_{2})\right\} \vert0\rangle^{+}
\end{equation}
with $K_{2}^{-+}$ being symmetric. The condition $a_{-}(k)\vert0\rangle^{-}=0$
leads to
\begin{equation}
0=\int_{0}^{\infty}k\tilde{dk'}\frac{1}{\omega(k)-\omega(k')}K_{1}^{-+}(k')\label{eq:K1-+}
\end{equation}
and 
\begin{equation}
\frac{1}{\omega(k)+\omega(k')}-\int_{0}^{\infty}\tilde{dk_{1}}\frac{1}{\omega(k_{1})-\omega(k)}K_{2}^{-+}(k_{1},k')=0\label{eq:K2-+}
\end{equation}
Solving the equations (\ref{eq:K1+-}-\ref{eq:K2-+}) from scratch
is a demanding task, but we can still check that the prediction from
the bootstrap approach does satisfy them.

\subsubsection{Bootstrap predictions}

\begin{figure}
\begin{centering}
\includegraphics[width=8cm]{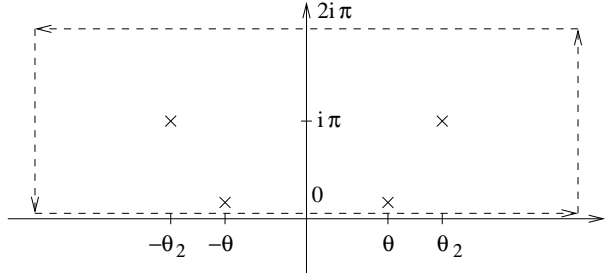}
\par\end{centering}

\protect\caption{Contour of integration and poles of the integrand for checking the
one particle term.}
\label{contour}
\end{figure}

Comparing the bosonic algebra (\ref{eq:N_D_mode_algebra}) to the
free boson Zamolodchikov-Faddeev algebra (\ref{sec:ZF_algebra}) shows
that they only differ in the normalization, $Z(\theta)=\frac{1}{\sqrt{2}}a(k)$,
with $k=m\sinh\theta$. Then we can relate the one-particle form factor
to the $K_{1}$ kernel, as
\begin{equation}
F_{1}^{-+}(\theta)=\frac{1}{\sqrt{2}}\phantom{I}^{-}\langle0\vert a_{+}^{+}(k)\vert0\rangle^{+}=\frac{1}{\sqrt{2}}\mathcal{N}K_{1}^{-+*}(k).
\end{equation}
From the bootstrap approach we get
\begin{equation}
F_{1}^{-+}(\theta)=\mathcal{N}2i\cosh\frac{\theta}{2}
\end{equation}
where $\mathcal{N}$ plays the role of the ground state expectation
value. To see that the resulting $K_{1}^{-+}$ kernel satisfy (\ref{eq:K1-+})
let us rewrite it in term of rapidity variables%
\footnote{To avoid the pole singularity on the real line we used the previous
$\epsilon$- prescription.%
},
\begin{equation}
\int_{0}^{\infty}\frac{d\theta'}{2\pi i}I_{1}(\theta'\vert\theta)=0\quad,\qquad I_{1}(\theta'\vert\theta)=\frac{1}{\cosh\theta'-\cosh\theta}\cosh\frac{\theta'}{2}.
\end{equation}
By observing that 
\begin{equation}
I_{1}(\theta'\vert\theta)=I_{1}(-\theta'\vert\theta)=-I_{1}(\theta'+2i\pi\vert\theta)=-I_{1}(-\theta'+2i\pi\vert\theta)
\end{equation}
we can extend the integration contour, depicted on Figure \ref{contour},
and get
\begin{equation}
\int_{0}^{\infty}\frac{d\theta'}{2\pi i}I_{1}(\theta'\vert\theta)=\frac{1}{4}\oint_{\mathcal{C}}\frac{d\theta'}{2\pi i}I_{1}(\theta'\vert\theta)=0
\end{equation}
where in the last step we applied the Residue theorem and the cancellation
of the residues.

Similarly, one can relate the $K_{1}^{+-}$ kernel to the one particle
form factor, as
\begin{equation}
F_{1}^{-+}(\theta+i\pi)=\frac{1}{\sqrt{2}}\langle-\vert a_{-}(k)\vert+\rangle=\frac{1}{\sqrt{2}}\mathcal{N}K_{1}^{+-}(k)
\end{equation}
Then, in the rapidity variables (\ref{eq:K1+-}) takes the form
\begin{equation}
\int_{0}^{\infty}\frac{d\theta'}{2\pi i}J_{1}(\theta'\vert\theta)=0\quad,\qquad J_{1}(\theta'\vert\theta)=\frac{\sinh\theta'\sinh\frac{\theta'}{2}}{\cosh\theta'-\cosh\theta}.
\end{equation}
Again, $J_{1}$ obeys the property
\begin{equation}
J_{1}(\theta'\vert\theta)=J_{1}(-\theta'\vert\theta)=-J_{1}(\theta'+2i\pi\vert\theta)=-J_{1}(-\theta'+2i\pi\vert\theta)
\end{equation}
Closing the contour as before and applying the Residue theorem proves
(\ref{eq:K1+-}).

In an analogous way one finds
\begin{equation}
F_{2}^{-+}(\theta_{1},\theta_{2})=\frac{1}{2}\mathcal{N}K_{2}^{-+*}(k_{1},k_{2})\quad,\qquad F_{2}^{-+}(\theta_{1}+i\pi,\theta_{2}+i\pi)=\frac{1}{2}\mathcal{N}K_{2}^{+-}(k_{1},k_{2})
\end{equation}
with the bootstrap solution of the form factor axioms given as
\begin{equation}
F_{2}^{-+}(\theta_{1},\theta_{2})=-4\mathcal{N}\frac{\cosh\frac{\theta_{1}}{2}\cosh\frac{\theta_{2}}{2}}{\cosh\theta_{1}+\cosh\theta_{2}}
\end{equation}
The equations (\ref{eq:K2-+}) and (\ref{eq:K2+-}) takes the form
\begin{eqnarray}
\frac{1}{\cosh\theta+\cosh\theta'} & = & -4i\int_{0}^{\infty}\frac{d\theta_{1}}{2\pi i}I_{2}(\theta_{1}\vert\theta,\theta')\nonumber \\
\frac{\sinh\theta'}{\cosh\theta+\cosh\theta'} & = & 4i\int_{0}^{\infty}\frac{d\theta_{1}}{2\pi i}J_{2}(\theta_{1}\vert\theta,\theta')\label{eq:K2_rewritten}
\end{eqnarray}
with
\begin{eqnarray}
I_{2}(\theta_{1}\vert\theta,\theta') & = & \frac{1}{\cosh\theta_{1}-\cosh\theta}\frac{\cosh\frac{\theta_{1}}{2}\cosh\frac{\theta'}{2}}{\cosh\theta_{1}+\cosh\theta'}\nonumber \\
J_{2}(\theta_{1}\vert\theta,\theta') & = & \frac{\sinh\theta_{1}}{\cosh\theta_{1}-\cosh\theta}\frac{\sinh\frac{\theta_{1}}{2}\sinh\frac{\theta'}{2}}{\cosh\theta_{1}+\cosh\theta'}
\end{eqnarray}
satisfying
\begin{alignat}{2}
I_{2}(\theta_{1}\vert\theta,\theta')= & I_{2}(-\theta_{1}\vert\theta,\theta')= & -I_{2}(\theta_{1}+2i\pi\vert\theta,\theta')= & -I_{2}(-\theta_{1}+2\pi i\vert\theta,\theta')\nonumber \\
J_{2}(\theta_{1}\vert\theta,\theta')= & J_{2}(-\theta_{1}\vert\theta,\theta')= & -J_{2}(\theta_{1}+2i\pi\vert\theta,\theta')= & -J_{2}(-\theta_{1}+2\pi i\vert\theta,\theta')
\end{alignat}
so that we can again close the contour as depicted on Figure \ref{contour}.
Applying the Residue theorem then proves (\ref{eq:K2_rewritten}). 

To summarize, the predictions of the bootstrap approach,
\begin{eqnarray}
K_{1}^{-+}(k)=-i2\sqrt{2}\cosh\frac{\theta}{2} & \quad,\qquad & K_{2}^{-+}(k_{1},k_{2})=-8\frac{\cosh\frac{\theta_{1}}{2}\cosh\frac{\theta_{2}}{2}}{\cosh\theta_{1}+\cosh\theta_{2}}\nonumber \\
K_{1}^{+-}(k)=-2\sqrt{2}\sinh\frac{\theta}{2} & \quad,\qquad & K_{2}^{+-}(k_{1},k_{2})=-8\frac{\sinh\frac{\theta_{1}}{2}\sinh\frac{\theta_{2}}{2}}{\cosh\theta_{1}+\cosh\theta_{2}}
\end{eqnarray}
does satisfy the constraints derived directly in the field theoretical
approach and thus provides an explicit relation between the incoming
and outgoing vacua, up to an overall normalization.

\subsection{Direct finite volume calculation}

In this subsection we map our problem to the open/closed string vertex
problem. In doing so we put the system in finite volume by introducing
another boundary at $x=-L$ with Neumann boundary condition. Eventually
we will take the limit $L\to\infty$ , thus the boundary condition
at $x=-L$ is irrelevant. 

If the right boundary at $x=0$ is chosen to be Neumann then the complete
system, satisfying the equations of motion and the boundary conditions,
is given as
\begin{equation}
f_{2n}^{+}(x)=\begin{cases}
\sqrt{\frac{2}{L}}\cos(k_{2n}x) & \quad n\in\mathbb{Z}^{+}\\
\frac{1}{\sqrt{L}} & \quad n=0
\end{cases}\qquad;\qquad k_{2n}=2n\frac{\pi}{2L}.
\end{equation}
Have we chosen the right boundary to be Dirichlet, we would get the
complete system
\begin{equation}
f_{2m+1}^{-}(x)=\sqrt{\frac{2}{L}}\sin(k_{2m+1}x)\quad,\quad m\in\mathbb{Z}_{0}^{+}\qquad;\qquad k_{2m+1}=(2m+1)\frac{\pi}{2L}.
\end{equation}
Thus for $t<0$ we have even, while for $t>0$ we have odd modes and
they never coincide. They are normalized as
\begin{equation}
\langle f_{2n}^{+}\vert f_{2n'}^{+}\rangle=\delta_{nn'}\qquad;\qquad\langle f_{2m+1}^{-}\vert f_{2m'+1}^{-}\rangle=\delta_{mm'}
\end{equation}
and they form separately a complete set
\begin{equation}
\sum_{n=0}^{\infty}f_{2n}(x)f_{2n}(y)=\delta(x-y)\quad,\quad\sum_{m=0}^{\infty}f_{2m+1}(x)f_{2m+1}(y)=\delta(x-y)\quad,\quad x,y\in\left[-L,0\right]\label{eq:Completeness_FinVol}
\end{equation}
where we introduced the scalar product $\langle f\vert g\rangle=\int_{-L}^{0}f(x)g(x)dx$.
Their overlaps are
\[
\langle f_{2m+1}^{-}\vert f_{2n}^{+}\rangle\equiv A^{-+}(2m+1,2n)=\langle f_{2n}^{+}\vert f_{2m+1}^{-}\rangle\equiv A^{+-}(2n,2m+1)=\begin{cases}
\frac{\sqrt{2}}{L}\frac{k_{2m+1}}{\omega_{0}^{2}-\omega_{2m+1}^{2}} & n=0\\
\frac{2}{L}\frac{k_{2m+1}}{\omega_{2n}^{2}-\omega_{2m+1}^{2}} & n\geq1
\end{cases}
\]
The field obeys the mode expansion
\begin{eqnarray}
\Phi(x,t)\hspace{-0.3cm} & =\hspace{-0.3cm} & \begin{cases}
\sum_{n=0}^{\infty}\frac{f_{2n}^{+}(x)}{\sqrt{2}\omega_{2n}}\left(a_{+}(2n)e^{-i\omega_{2n}t}+a_{+}^{+}(2n)e^{i\omega_{2n}t}\right) & t<0\\
\sum_{m=0}^{\infty}\frac{f_{2m+1}^{-}(x)}{\sqrt{2}\omega_{2m+1}}\left(a_{-}(2m+1)e^{-i\omega_{2m+1}t}+a_{-}^{+}(2m+1)e^{i\omega_{2m+1}t}\right) & t>0
\end{cases}\label{eq:Field_FinVol}
\end{eqnarray}
with $\omega_{n}=\sqrt{m^{2}+k_{n}^{2}}$. The commutation relations
turns out to be 
\begin{equation}
\left[a_{+}(2n),a_{+}^{+}(2m)\right]=\omega_{2n}\delta_{nm}\quad,\qquad\left[a_{-}(2n+1),a_{-}^{+}(2m+1)\right]=\omega_{2n+1}\delta_{nm}\label{eq:algebra_FinVol}
\end{equation}
Similarly to the infinite volume case we can relate the modes to each
other by demanding the continuity the field and its momentum $\Pi=\partial_{t}\Phi$.
Projecting $\Phi(x,t=0)$ and $\Pi(x,t=0)$ onto $\langle f_{n}^{\pm}\vert$
and combining them results in
\begin{multline}
a_{+}(2n)=\sum_{m=0}^{\infty}\frac{A^{+-}(2n,2m+1)}{2\omega_{2m+1}}\left\{ (\omega_{2n}+\omega_{2m+1})a_{-}(2m+1)+(\omega_{2n}-\omega_{2m+1})a_{-}^{+}(2m+1)\right\} \\
\hspace{-0.1cm}a_{-}(2m+1)=\sum_{n=0}^{\infty}\frac{A^{-+}(2m+1,2n)}{2\omega_{2n}}\left\{ (\omega_{2m+1}+\omega_{2n})a_{+}(2n)+\left(\omega_{2m+1}-\omega_{2n}\right)a_{+}^{+}(2n)\right\} \label{eq:a_N_D}
\end{multline}
and the conjugate relations. The compatibility of these relations
is granted due to the unitarity 
\begin{eqnarray}
\sum_{m=0}^{\infty}A^{+-}(2n,2m+1)A^{-+}(2m+1,2n_{1}) & = & \delta_{n,n_{1}}\nonumber \\
\sum_{n=0}^{\infty}A^{-+}(2m+1,2n)A^{+-}(2n,2m_{1}+1) & = & \delta_{m,m_{1}}\label{eq:Unitarity_boson}
\end{eqnarray}

The states are built over a Fock vacuum defined as
\begin{equation}
a_{+}(2n)\vert0\rangle^{+}=0\quad,\qquad a_{-}(2n+1)\vert0\rangle^{-}=0\qquad n=0,1,2,\dots
\end{equation}
A multiparticle Neumann/Dirichlet state are generated by repeated
action of creation operators on the corresponding vacuum state, 
\begin{eqnarray}
\vert\{n_{1},\dots,n_{N}\}\rangle^{+} & = & a_{+}^{+}(2n_{1})\dots a_{+}^{+}(2n_{N})\vert0\rangle^{+}\nonumber \\
\vert\{m_{1},\dots,m_{M}\}\rangle^{-} & = & a_{-}^{+}(2m_{1}+1)\dots a_{-}^{+}(2m_{M}+1)\vert0\rangle^{-}
\end{eqnarray}

We are interested in the overlap of an incoming Neumann state and
an outgoing Dirichlet state. To this end let us flip back the outgoing
Dirichlet states to independent incoming ones by hermitian conjugation.
To distinguish the flipped states from the original ones let us introduce
a new set of bosonic operators as
\begin{equation}
a_{+}(2n)\mapsto c_{1}(2n)\qquad,\qquad a_{-}(2m+1)\mapsto c_{2}^{+}(2m+1)
\end{equation}
As the hermitian conjugation reverses the order of the operators,
the new ones satisfy the algebra
\begin{equation}
\left[c_{1}(2n),c_{1}^{+}(2n')\right]=\omega_{2n}\delta_{nn'}\quad,\qquad\left[c_{2}(2m+1),c_{2}^{+}(2m'+1)\right]=\omega_{2m+1}\delta_{m,m'}
\end{equation}
and the other commutators vanish. An incoming state is now built over
the Fock vacuum defined as
\begin{equation}
c_{1}(2n)\vert0,0\rangle=0\quad,\qquad c_{2}(2m+1)\vert0,0\rangle=0
\end{equation}
and the states are generated by repeated action of creation operators
\begin{equation}
\hspace{-0.2cm}\vert\{n_{1},\dots,n_{N}\},\{m_{1},\dots,m_{M}\}\rangle=c_{1}^{+}(2n_{1})\dots c_{1}^{+}(2n_{N})c_{2}^{+}(2m_{1}+1)\dots c_{2}^{+}(2m_{M}+1)\vert0,0\rangle
\end{equation}
We define the vertex state $\vert V\rangle$ as
\begin{equation}
\phantom{I}^{-}\langle\{m_{1},\dots,m_{M}\}\vert\{n_{1},\dots,n_{N}\}\rangle^{+}\equiv\langle V\vert\{n_{1},\dots,n_{N}\},\{m_{1},\dots,m_{M}\}\rangle
\end{equation}
We parametrize it as
\begin{equation}
\vert V\rangle=\mathcal{N}^{*}e^{\Delta}\vert0,0\rangle
\end{equation}
with
\begin{eqnarray}
\Delta & = & \sum_{n_{1},n_{2}=0}^{\infty}\frac{V_{++}(2n_{1},2n_{2})}{2}\frac{c_{1}^{+}(2n_{1})c_{1}^{+}(2n_{2})}{\omega_{2n_{1}}\omega_{2n_{2}}}+\sum_{n,m=0}^{\infty}V_{+-}(2n,2m+1)\frac{c_{1}^{+}(2n)c_{2}^{+}(2m+1)}{\omega_{2n}\omega_{2m+1}}+\nonumber \\
 &  & +\sum_{m_{1},m_{2}=0}^{\infty}\frac{V_{--}(2m_{1}+1,2m_{2}+1)}{2}\frac{c_{2}^{+}(2m_{1}+1)c_{2}^{+}(2m_{2}+1)}{\omega_{2m_{1}+1}\omega_{2m_{2}+1}}
\end{eqnarray}
and $\mathcal{N}=\phantom{I}^{-}\langle0\vert0\rangle^{+}$. The $V_{\pm\pm}$
functions are called the Neumann coefficients, $V_{++}$ and $V_{--}$
are symmetric in their arguments.

After flipping, the relations (\ref{eq:a_N_D}) become
\begin{multline}
c_{1}(2n)-\sum_{m=0}^{\infty}\frac{A^{+-}(2n,2m+1)}{2\omega_{2m+1}}\left\{ (\omega_{2n}-\omega_{2m+1})c_{2}(2m+1)+(\omega_{2n}+\omega_{2m+1})c_{2}^{+}(2m+1)\right\} =0\\
c_{2}(2m+1)-\sum_{n=0}^{\infty}\frac{A^{-+}(2m+1,2n)}{2\omega_{2n}}\left\{ -(\omega_{2n}-\omega_{2m+1})c_{1}(2n)+(\omega_{2n}+\omega_{2m+1})c_{1}^{+}(2n)\right\} =0\label{eq:Boson_relations}
\end{multline}
and their hermitian conjugates, where the equations are understood
in the weak sense, i.e. when sandwiched between the vertex state and
any multiparticle state. These relations constrain the Neumann coefficients,
resulting an overdetermined, nevertheless consistent system of equations.
The only three independent ones are
\begin{eqnarray}
\delta_{n,n_{1}}-\frac{1}{\omega_{2n_{1}}}\sum_{m=0}^{\infty}\frac{\omega_{2n}+\omega_{2m+1}}{2\omega_{2m+1}}A^{+-}(2n,2m+1)V_{+-}^{*}(2n_{1},2m+1) & = & 0\nonumber \\
\sum_{m_{1}=0}^{\infty}\frac{\omega_{2n}+\omega_{2m_{1}+1}}{\omega_{2m_{1}+1}}A^{+-}(2n,2m_{1}+1)V_{--}^{*}(2m+1,2m_{1}+1)+\qquad\nonumber \\
+\left(\omega_{2n}-\omega_{2m+1}\right)A^{+-}(2n,2m+1) & = & 0\nonumber \\
V_{++}^{*}(2n,2n_{1})-\sum_{m=0}^{\infty}\frac{\omega_{2n}-\omega_{2m+1}}{2\omega_{2m+1}}A^{+-}(2n,2m+1)V_{+-}^{*}(2n_{1},2m+1) & = & 0\label{eq:FinVol_vertex}
\end{eqnarray}

A similar problem was analyzed in the context of the open-closed string
vertex \cite{Lucietti:2003ki,Lucietti:2004wy}. In the case when we
choose Dirichlet boundary condition on the open string the resulting
equations (eqs. (2.34-2.36) of \cite{Lucietti:2003ki}) can be mapped
to (\ref{eq:Boson_relations}), thus we can simply read of the Neumann
coefficients. Their volume dependence is encoded into some complicated
modified gamma functions which, however, take a relatively simple
form in the large volume limit. The large volume asymptotic solution
reads as
\begin{eqnarray}
V_{+-}(2n,2m+1)\hspace{-0.3cm} & =\hspace{-0.3cm} & \frac{1}{L}\frac{k_{2m+1}}{\omega_{2n}-\omega_{2m+1}}\frac{\omega_{2m+1}+\omega_{1}}{\omega_{2n}+\omega_{1}}\frac{(\omega_{2n}+M)^{3/2}}{(\omega_{2m+1}+M)^{3/2}}+O(e^{-ML})\nonumber \\
V_{--}(2m_{1}+1,2m_{2}+1)\hspace{-0.3cm} & =\hspace{-0.3cm} & \frac{1}{L}\frac{k_{2m_{1}+1}k_{2m_{2}+1}}{\omega_{2m_{1}+1}+\omega_{2m_{2}+1}}\frac{\omega_{2m_{1}+1}+\omega_{1}}{(\omega_{2m_{1}+1}+M)^{3/2}}\frac{\omega_{2m_{2}+1}+\omega_{1}}{(\omega_{2m_{2}+1}+M)^{3/2}}+O(e^{-ML})\nonumber \\
V_{++}(2n_{1},2n_{2})\hspace{-0.3cm} & =\hspace{-0.3cm} & -\frac{1}{L}\frac{1}{\omega_{2n_{1}}+\omega_{2n_{2}}}\frac{(\omega_{2n_{1}}+M)^{3/2}}{\omega_{2n_{1}}+\omega_{1}}\frac{(\omega_{2n_{2}}+M)^{3/2}}{\omega_{2n_{2}}+\omega_{1}}+O(e^{-ML})
\end{eqnarray}
where $M$ is the mass of the particles. The unnormalized two-particle
finite volume form factors are related to the Neumann coefficients
as
\begin{eqnarray}
\langle-\vert a_{+}^{+}(2n_{1})a_{+}^{+}(2n_{2})\vert+\rangle & = & \mathcal{N}V_{++}^{*}(2n_{1},2n_{2})\nonumber \\
\langle-\vert a_{-}(2m+1)a_{+}^{+}(2n)\vert+\rangle & = & \mathcal{N}V_{+-}^{*}(2n,2m+1)\nonumber \\
\langle-\vert a_{-}(2m_{1}+1)a_{-}(2m_{2}+1)\vert+\rangle & = & \mathcal{N}V_{--}^{*}(2m_{1}+1,2m_{2}+1)
\end{eqnarray}

We would like to take a sensible infinite volume limit $L\rightarrow\infty$,
while keeping the momenta fixed, $k_{2n}=k$ and $k_{2m+1}=k'$. The
dispersion relation does not change, $\omega_{2n}=\omega(k)$ and
$\omega_{2m+1}=\omega(k')$. Comparing the completeness relations
(\ref{eq:Completeness},\ref{eq:Completeness_FinVol}), the mode decomposition
of the field (\ref{eq:Field},\ref{eq:Field_FinVol}) and the algebra
relations (\ref{eq:N_D_mode_algebra},\ref{eq:algebra_FinVol}) in
finite and in infinite volume, one can read off the correct scaling
of the mode operators,
\begin{alignat}{4}
\sqrt{4L}a_{+}(2n) & \rightarrow & a_{+}(k) & \qquad,\qquad & \sqrt{4L}a_{+}^{+}(2n) & \rightarrow & a_{+}^{+}(k)\nonumber \\
-i\sqrt{4L}a_{-}(2m+1) & \rightarrow & a_{-}(k') & \qquad,\qquad & i\sqrt{4L}a_{-}^{+}(2m+1) & \rightarrow & a_{-}^{+}(k')
\end{alignat}
In this infinite volume limit $\omega_{1}\rightarrow M$, thus one
gets
\begin{eqnarray}
\hspace{-0.3cm}\phantom{I}^{-}\langle0\vert a_{+}^{+}(k_{1})a_{+}^{+}(k_{2})\vert0\rangle^{+}\hspace{-0.3cm} & = & \hspace{-0.3cm}-\mathcal{N}4\frac{\sqrt{\omega(k_{1})+M}\sqrt{\omega(k_{2})+M}}{\omega(k_{1})+\omega(k_{2})}=-\mathcal{N}8\frac{\cosh\frac{\theta_{1}}{2}\cosh\frac{\theta_{2}}{2}}{\cosh\theta_{1}+\cosh\theta_{2}}\nonumber \\
\hspace{-0.3cm}\phantom{I}^{-}\langle0\vert a_{-}(k')a_{+}^{+}(k)\vert0\rangle^{+}\hspace{-0.3cm} & = & \hspace{-0.3cm}-\mathcal{N}4i\frac{k'}{\omega(k)-\omega(k')}\frac{\sqrt{\omega(k)+M}}{\sqrt{\omega(k')+M}}=-\mathcal{N}8i\frac{\cosh\frac{\theta}{2}\sinh\frac{\theta'}{2}}{\cosh\theta-\cosh\theta'}\nonumber \\
\hspace{-0.3cm}\phantom{I}^{-}\langle0\vert a_{-}(k'_{1})a_{-}(k'_{2})\vert0\rangle^{+}\hspace{-0.3cm} & = & \hspace{-0.3cm}-\mathcal{N}4\frac{k'_{1}\, k'_{2}}{\omega(k'_{1})+\omega(k'_{2})}\frac{1}{\sqrt{\omega(k'_{1})+M}\sqrt{\omega(k'_{2})+M}}=\nonumber \\
 & = & \hspace{-0.3cm}-\mathcal{N}8\frac{\sinh\frac{\theta'_{1}}{2}\sinh\frac{\theta'_{2}}{2}}{\cosh\theta'_{1}+\cosh\theta'_{2}}
\end{eqnarray}

Comparing the bosonic algebra (\ref{eq:N_D_mode_algebra}) to the
free boson Zamolodchikov-Faddeev algebra (\ref{eq:bulk_ZF}) shows
that they only differ in the normalization, $Z(\theta)=\frac{1}{\sqrt{2}}a(k)$,
thus the form factors are
\begin{eqnarray}
\hspace{-0.3cm}F_{2}^{-+}(\theta_{1},\theta_{2})\hspace{-0.3cm} & = & \hspace{-0.5cm}\phantom{I}^{-}\langle0\vert Z^{+}(\theta_{1})Z^{+}(\theta_{2})\vert0\rangle^{+}=\frac{1}{2}\hspace{-0.1cm}\phantom{I}^{-}\langle0\vert a_{+}^{+}(k_{1})a_{+}^{+}(k_{2})\vert0\rangle^{+}\\
\hspace{-0.3cm}F_{2}^{-+}(i\pi+\theta',\theta)\hspace{-0.3cm} & = & \hspace{-0.5cm}\phantom{I}^{-}\langle0\vert Z^{+}(i\pi+\theta')Z^{+}(\theta)\vert0\rangle^{+}=\frac{1}{2}\hspace{-0.1cm}\phantom{I}^{-}\langle0\vert a_{-}(k')a_{+}^{+}(k)\vert0\rangle^{+}\nonumber \\
\hspace{-0.3cm}F_{2}^{-+}(i\pi+\theta'_{1},i\pi+\theta'_{2})\hspace{-0.3cm} & = & \hspace{-0.5cm}\phantom{I}^{-}\langle0\vert Z^{+}(i\pi+\theta'_{1})Z^{+}(i\pi+\theta'_{2})\vert0\rangle^{+}=\frac{1}{2}\hspace{-0.1cm}\phantom{I}^{-}\langle0\vert a_{-}(k'_{1})a_{-}(k'_{2})\vert0\rangle^{+}\nonumber 
\end{eqnarray}
As $\mathcal{N}$ plays the role of the vacuum expectation value,
this result coincides with the solutions of the bootstrap axioms.

\end{document}